\documentclass[aip,jcp,reprint,showpacs]{revtex4-1}
\usepackage{graphicx}
\usepackage{latexsym}
\usepackage{amsmath}
\usepackage{amssymb}
\usepackage{amsfonts}
\usepackage{mathrsfs}
\usepackage{color}
\usepackage{verbatim}

\def\br{\mathbf{r}}

\def\half{\frac{1}{2}}

\def\h2o{\mathrm{H}_2\mathrm{O}}

\begin{document}

\title{A systematic benchmark of the \textit{ab initio} Bethe-Salpeter equation approach for low-lying optical excitations of small organic molecules}

\author{Fabien Bruneval}
\affiliation{CEA, DEN, Service de Recherches de M\'etallurgie Physique, F-91191 Gif-sur-Yvette, France}
\affiliation{Molecular Foundry, Lawrence Berkeley National Laboratory, Berkeley, California 94720, United States}
\affiliation{Department of Physics, University of California, Berkeley, California 94720, United States}

\author{Samia M. Hamed}
\affiliation{Molecular Foundry, Lawrence Berkeley National Laboratory, Berkeley, California 94720, United States}
\affiliation{Department of Chemistry, University of California, Berkeley, California 94720, United States}
\affiliation{Department of Physics, University of California, Berkeley, California 94720, United States}
\affiliation{Kavli Energy Nanosciences Institute at Berkeley, Berkeley, California 94720, United States}

\author{Jeffrey B. Neaton}
\affiliation{Molecular Foundry, Lawrence Berkeley National Laboratory, Berkeley, California 94720, United States}
\affiliation{Department of Physics, University of California, Berkeley, California 94720, United States}
\affiliation{Kavli Energy Nanosciences Institute at Berkeley, Berkeley, California 94720, United States}

\begin{abstract}
The predictive power of the \textit{ab initio} Bethe-Salpeter equation (BSE) approach, rigorously based on many-body Green's function theory but incorporating information from density functional theory, has already been demonstrated for the optical gaps and spectra of solid-state systems. Interest in photoactive hybrid organic/inorganic systems has recently increased, and so has the use of the BSE for computing neutral excitations of organic molecules. However, no systematic benchmarks of the BSE for neutral electronic excitations of organic molecules exist. Here, we study the performance of the BSE for the 28 small molecules in Thiel's widely-used time-dependent density functional theory benchmark set [M. Schreiber \textit{et al}. J. Chem. Phys. \textbf{128}, 134110 (2008)]. We observe that the BSE produces results that depend critically on the mean-field starting point employed in the perturbative approach. We find that this starting point dependence is mainly introduced through the quasiparticle energies obtained at the intermediate $GW$ step, and that with a judicious choice of starting mean-field, singlet excitation energies obtained from BSE are in excellent quantitative agreement with higher-level wavefunction methods. The quality of the triplet excitations is slightly less satisfactory.
\end{abstract}
\maketitle

\section{Introduction}

Optical properties are of broad fundamental and practical interest. For example, they determine the color of everyday objects; they dictate the absorption and transfer of 
photons by and between chromophores embedded in protein environments; and they control the fluorescence behavior of molecules used as markers in biomedical imaging applications. These diverse phenomena (and many others) are in fact united by the same underlying quantum mechanics describing electronic excitations and their consequences.

The ability to reliably and quantitatively predict these neutral excitations from computer calculations is an important goal, and several competing methods for this purpose are in use today. Besides the wavefunction-based methods which are computationally-intensive and thus often limited to relatively small systems, two main formalisms are present in the literature: time-dependent density functional theory (TD-DFT)\cite{runge_prl1984,marques_arpc2004} and the Bethe-Salpeter equation (BSE)\cite{strinati_rnc1988} approach from many-body perturbation theory. Both approaches have been widely used but for different classes of systems: TD-DFT is primarily for molecules, and BSE for solids.\cite{onida_rmp2002} 

TD-DFT has thus far been applied with great success to the calculations of the low-lying excitations of isolated molecules. It is a computationally attractive method that can be used efficiently even for relatively large systems, and performs particularly well when paired with a hybrid exchange correlation (xc) functional.\cite{bauernschmitt_cpl1996} However, TD-DFT, with standard hybrids, can be challenged by Rydberg final states and charge transfer excitations.\cite{tozer_jcp1998,dreuw_jacs2004} These failures can be tempered by employing xc approximations that contain the full long-range exact exchange contribution, as do the tuned-CAM-BL3YP,\cite{okuno_jppa2012} BNL,\cite{baer_prl2005,livshits_pccp2007} and OT-RSH functionals.\cite{refaely_prl2012} However, and despite interesting proposals \cite{sottile_prl2003,marini_prl2003,sharma_prl2011}, for solids no purely \textit{ab initio} approximation to TD-DFT is available to date, since the content of long-range exact exchange in solids should be modulated by the system-dependent dielectric constant.\cite{botti_rpp2007,bruneval_jcp2006}


Likewise the BSE, whose solution is the two-particle electron-hole Green's function, also accounts for the screened electron-hole interaction; the BSE framework has been shown to be extremely successful in predicting the optical spectra of bulk solids\cite{hanke_prb1975,albrecht_prl1998,shirley_prl1998,rohlfing_prl1998} and of low-dimensional materials.\cite{spataru_prl2004} In part inspired by recent interest in organic-based energy conversion materials, BSE has begun to be applied to finite organic molecular systems as well.\cite{blase_prb2011,faber_prb2011,sharifzadeh_prb2012,sharifzadeh_jpcl2013,rebolini_jcp2014,rocca_jctc2014} Relative to standard contemporary TD-DFT approaches, the BSE method has many attractive features: through the \textit{ab initio} calculation of the screened Coulomb interaction, the electron-hole interaction has the correct asymptotic behavior independent of the system, be it a bulk solid, a low-dimensional nanostructure or polymer, or a molecule. This feature results, for instance, in a correct description of charge transfer excitations in molecules.\cite{blase_apl2011}

Additionally, the description of neutral excitations within the BSE are built upon a foundation of charged excitation energies, corresponding to electron addition or removal, determined within the $GW$ approximation\cite{hedin_pr1965,strinati_rnc1988,aulbur_review2000}. The $GW$ approach is known to yield much more accurate values of the fundamental (or quasiparticle) gap energy for a variety of systems than, e.g., DFT. This is in contrast with TD-DFT for which underlying Kohn-Sham eigenvalues have little physical meaning.\cite{parr_book}
Only the highest occupied molecular orbital (HOMO) and the lowest unoccupied molecular orbital (LUMO) energies can be safely interpreted as the negative of ionization potential (IP) and the negative of electron affinity (EA) in a generalized Kohn-Sham scheme (gKS) (including hybrid functionals).\cite{cohen_science2008,cohen_prb2008,yang_jcp2012}
All the other eigenvalues are not, strictly speaking, observables, although recent work on tuned range-separated hybrids suggests that both quasiparticle gaps and outer valence spectra from a gKS approach can be in quantitative agreement with photoemission and $GW$ calculations.
\cite{kronik_jctc2012, refaely_prl2012,egger_jctc2014}

For these reasons, the BSE approach is increasingly being used to predict excitation energies for molecules and is an alternative to TD-DFT. However, there are, to date, no general assessments of the quality of BSE results for low-lying neutral excitations of isolated molecules. Although several benchmarks of TD-DFT have been reported detailing its accuracy for different choices of xc functional relative to wavefunction-based methods,\cite{laurent_ijqc2013} no such systematic effort has been undertaken for the BSE approach. Here, we evaluate the accuracy of BSE neutral excitations compared to values for 28 small organic molecules calculated by Thiel and coworkers with high-level wavefunction-based methods. \cite{schreiber_jcp2008,silvajunior_jcp2008} This set of 28 molecules, hereafter referred to as ``Thiel's set'', includes 103 singlet and 63 triplet excitations, all computed with multiple coupled-cluster level methods.\cite{bartlett_rmp2007} Following prior studies with TD-DFT, we benchmark to theoretical values, rather than experimental data: the compared calculations employ the same basis set, all atomic positions are identical, vibrations and temperature effects are neglected, and there is no solvent or other environmental conditions. With this benchmark, we are able to provide a general assessment, as well as guidelines and rationale for the successful application of BSE to molecular systems.

The article is organized as follows:
Sec.~II is a general presentation of the BSE formalism.
Sec.~III details the practical calculations and presents Thiel's set.
Sec.~IV presents the BSE results for the Thiel's set.
In Sec.~V we discuss the BSE results and analyses the causes of success and of failures.
The study is concluded in Sec.~VI.
Hartree atomic units will be used throughout the text ($\hbar = e = a_0 = 1$).

\section{Bethe-Salpeter equation for molecules}
\label{secII}

\begin{figure}[t]
\includegraphics[width=\columnwidth]{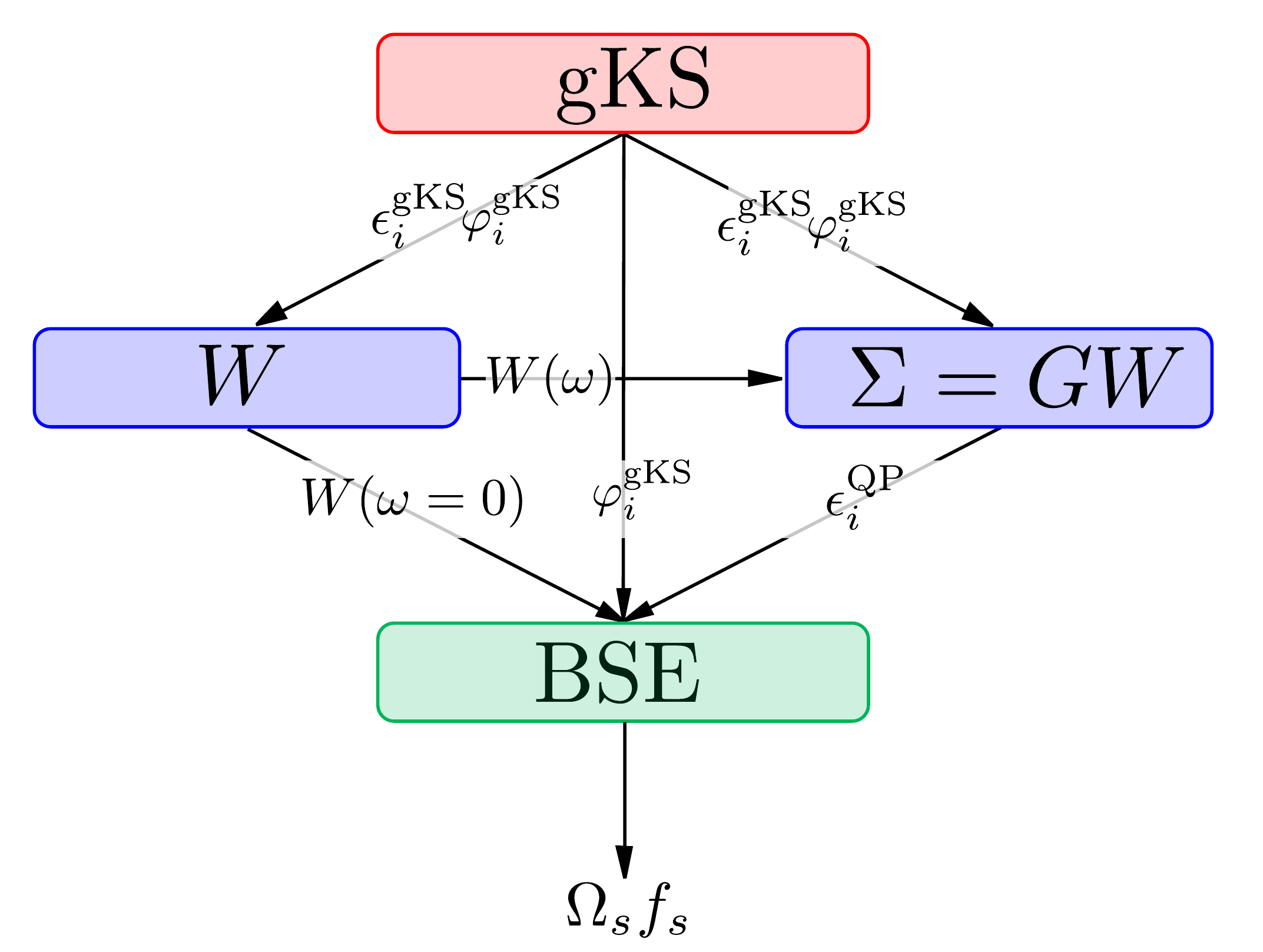}
\caption{Typical BSE calculation flowchart.
\label{fig:bse_flowchart}
}
\end{figure}

The BSE is an equation for the two-particle Green's function, more precisely for its electron-hole time-ordering.
Rigorously derived from many-body perturbation theory, the BSE, when using a static approximation,
is completely analogous to TD-DFT expressed through Casida's equations.\cite{casida_book1995}
Here we describe the most salient features of a typical BSE solution in its most common practical implementation.\cite{onida_rmp2002}

Figure~\ref{fig:bse_flowchart} shows a flowchart of a BSE evaluation of neutral excitation energies.
The calculation consists of 3 main steps:
a self-consistent mean-field DFT calculation in the gKS scheme;
a perturbative $GW$ calculation to obtain the quasiparticle energies;
and a final BSE solution to produce the excitation energies $\Omega_s$ and the corresponding oscillator strengths $f_s$.

The initial DFT step is denoted gKS since the starting mean-field might be standard Kohn-Sham with a local xc approximation,
or it might be derived from a more general hybrid functional or even Hartree-Fock.
This first step produces eigenvalues and wavefunctions that are used to evaluate the screened Coulomb interaction $W$
and the $GW$ self-energy.

Details of the calculation of the $GW$ self-energy for atoms and molecules can be found in Ref.~\onlinecite{bruneval_jcp2012}.
Let us simply write the self-energy $\Sigma$ as
\begin{equation}
 \Sigma(1,2) = i G(1,2) W(1^+,2),
\end{equation}
where the composite index 1 is short for position, time, and spin $(\br_1,t_1,\sigma_1)$.
$1^+$ indicates the limit as time goes to $t_1$ from above.
$G$ is the one-particle Green's function and $W$ is the screened Coulomb interaction. 
The $GW$ self-energy produces quasiparticle energies which are, by definition, the binding energies of electrons or holes in a system. These energies are precisely the observables measured by photoemission (for occupied states) and inverse-photoemission (for unoccupied states).
In practice, the $GW$ quasiparticle energies show good agreement with experiment, albeit with a notable starting point dependence. We will discuss this further below.
\cite{grossman_prl2001,blase_prb2011,ren_njp2012,korzdorfer_prb2012,sharifzadeh_epjb2012,bruneval_jctc2013,koval_prb2014,korbel_jctc2014}

In the end, BSE is a Dyson-like equation for the so-called two-particle correlation function $L$.
The full equation reads
\begin{multline}
 \label{eq:bse}
 L( 1,2;1',2') = L_0(1,2; 1',2') \\
   + \int d3 d4 d5 d6 L_0(1,2;4,3) \\
       \times \frac{\delta M(3,4)}{\delta G(6,5)} L(6,2;5,2'),
\end{multline}
where the non-interacting correlation function $L_0$ is expressed as
\begin{equation}
  L_0(1,2; 1',2') = G(1,2') G(2,1')
\end{equation}
and $M$ is simply the sum of the Hartree potential and the self-energy
\begin{equation}
  M(3,4) =  v_H(3)\delta(3,4) + \Sigma(3,4) . 
\end{equation}
When the indices are contracted, $L$ and $L_0$ yield the usual interacting and non-interacting polarizabilities
\begin{subequations}
\begin{eqnarray}
  \chi(1,2)   & =&  -i L(1,2;1^+,2^+)  \\
  \chi_0(1,2) & =&  -i L_0(1,2;1^+,2^+) .
\end{eqnarray}
\end{subequations}
When expressed in this form, the BSE in Eq.~(\ref{eq:bse}) and the central equations of TD-DFT in the linear response formalism
\begin{equation}
 \chi(1,2) = \chi_0(1,2) + \int d3 d4 \chi_0(1,3) \frac{\delta v_\text{KS}(3) } {\delta \rho(4)} \chi(4,2)
\end{equation}
are linked in rather intuitive fashion.

In practice, the BSE is generally solved using the screened Hartree-Fock approximation to $\Sigma$, a choice that can alternatively be viewed as a $GW$ approximation to $\Sigma$ in the static limit. Hence, the BSE kernel simplifies to the following frequency independent expression:
\begin{multline}
 \frac{\delta M(\br_3,\br_4)}{\delta G(\br_6,\br_5)} =
              -i v(\br_3-\br_5) \delta(\br_3-\br_4)\delta(\br_5-\br_6)  \\
                + i W(\br_3,\br_4,\omega=0) \delta(\br_3-\br_5)\delta(\br_4-\br_6) ,
\end{multline}
where $v$ is the bare Coulomb interaction in the previous equation.

With this static assumption, the BSE can be recast into a matrix form in a transition space spanned by the orbital products $\varphi_i(\br)\varphi_j(\br)$ where pairs of states $i$ and $j$ are either occupied/unoccupied or unoccupied/occupied. Thus the BSE results in an eigenvalue problem with the same block form as the TD-DFT equations
\begin{equation}
 \label{eq:bse_blocks}
 \left( 
   \begin{array}{cc}
      \phantom{-}A  &  \phantom{-}B \\
      -B  & -A 
   \end{array}
 \right) 
   \left(
   \begin{array}{c}
       X_s \\
       Y_s 
   \end{array}
   \right)
  = \Omega_s
   \left(
   \begin{array}{c}
       X_s \\
       Y_s 
   \end{array}
   \right) .
\end{equation}
where $\Omega_s$ are the neutral excitations and  $(X_s , Y_s)$ are the eigenvectors. The complex conjugation
has been dropped because the wavefunctions are assumed to be real-valued.
Just as in TD-DFT, the upper block $A$ accounts for resonant transitions from occupied to unoccupied orbitals, whereas
the lower block $-A$ accounts for the antiresonant transitions, and the two types of transitions are coupled through the blocks $B$ and $-B$. The neglect of the coupling $B$ leads to the Tamm-Dancoff approximation.\cite{ullrich_book2012}

The only difference between TD-DFT and BSE lies in the specific expression of the matrix elements in $A$ and $B$.
In BSE, if $i$ and $j$ are occupied states and $a$ and $b$ are unoccupied states, these elements read,
for spin-restricted calculations,
\begin{subequations}
\begin{eqnarray}
 \label{eq:bse_elementsa}
  A_{ia}^{jb} & =& (\epsilon_a^\text{QP} - \epsilon_i^\text{QP})
               \delta_{ij}\delta_{ab} \nonumber \\
          &  &  - \alpha^\text{S/T} ( i a | j b ) + W_{ij}^{ab}(\omega=0) \\
 \label{eq:bse_elementsb}
  B_{ia}^{jb} & =& -\alpha^\text{S/T}( i a | b j ) + W_{ib}^{aj}(\omega=0) ,
\end{eqnarray}
\end{subequations}
where $( i a | j b )$ are Coulomb integrals in Mulliken notation.
The coefficient $\alpha^\text{S/T}$ is set to 2 in the case of a singlet final state or to 0 in the case of a triplet final state.

The eigenvalue problem posed by the BSE as shown in Eq.~(\ref{eq:bse_blocks}) is numerically cumbersome:
the matrix size grows as the square of the number of atoms (2 times the number of occupied states times the number of unoccupied states) and
it is furthermore a non-symmetric eigenvalue problem.
However it is well known from TD-DFT that this problem can be reduced to a symmetric eigenvalue problem whose size is cut in half.\cite{casida_book1995,ullrich_book2012}
After some algebra, the problem can be recast as
\begin{equation}
  C Z_s = \Omega_s^2 Z_s ,
\end{equation}
where $C = ( A - B)^{1/2} (A+B) (A-B)^{1/2}$ is a symmetric matrix that is half the size of the initial problem in Eq.~(\ref{eq:bse_blocks}).
The above expression assumes matrix $(A-B)$ to be positive definite.
From the knowledge of an eigenvector $Z_s$, one can build both $X_s$ and $Y_s$ as
\begin{subequations}
\begin{eqnarray}
  X_s  &=&  \half \left[ (A-B)^{1/2} + \Omega_s (A-B)^{-1/2}  \right] Z_s  \\
  Y_s  &=&  \half \left[ (A-B)^{1/2} - \Omega_s (A-B)^{-1/2}  \right] Z_s .
\end{eqnarray}
\end{subequations}
In fact, here the calculation of the square root of matrix $(A-B)$ requires another diagonalization.
Note that in TD-DFT, $(A-B)$ is a diagonal matrix and its square root is readily obtained.
However recent work\cite{shao_arxiv2015} has proven that Cholesky decompositions can be a work-around to avoid this second diagonalization.

A TD-DFT calculation would be essentially identical to this except that the eigenvalues entering in $A$ would
be gKS eigenvalues instead of quasiparticle energies from a GW approxomation, and except that the $W$ term would be replaced by the xc kernel $f_{xc}$
(with a different index ordering).

Having reviewed the BSE formalism, we will turn to the practical application of it to Thiel's set.
\cite{schreiber_jcp2008,silvajunior_jcp2008}

\section{Thiel's set: technical aspects}
\label{secIII}

\begin{figure}[t]
\includegraphics[width=\columnwidth]{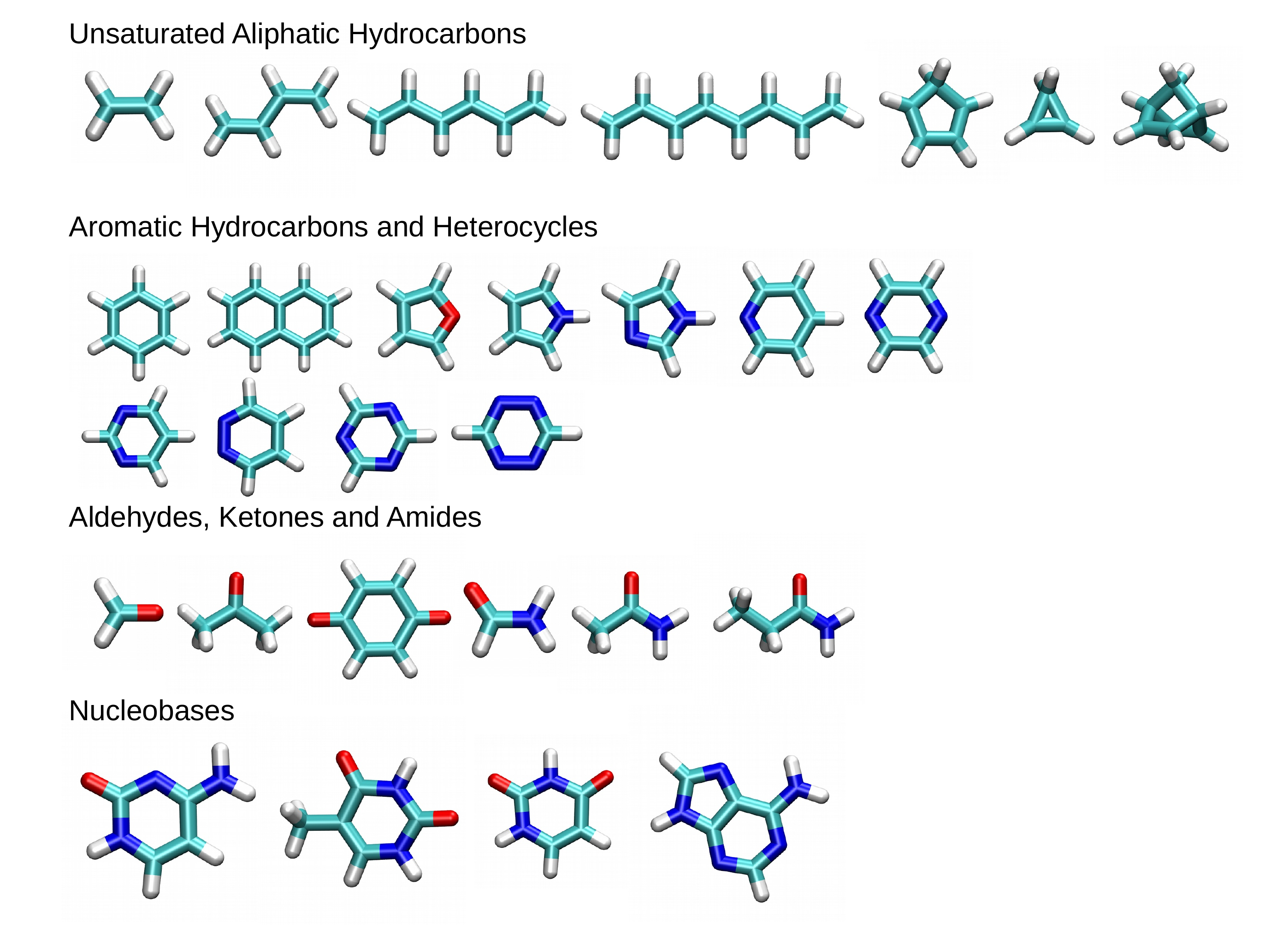}
\caption{The 28 molecules contained in Thiel's set.
H is white, C is light blue, N is dark blue, and O is red.
\label{fig:thiel_set_picture}
}
\end{figure}

In the present study we evaluate the quality of the BSE neutral excitation energies for Thiel's set of 28 small organic molecules.\cite{schreiber_jcp2008,silvajunior_jcp2008}

Our calculations are performed with the \textsc{molgw} code\cite{bruneval_jcp2012,molgw} which is an implementation
of $GW$ and BSE many-body perturbation theory with Gaussian basis functions.
\textsc{molgw} relies on an external library, \textsc{libint},\cite{libint2} to evaluate the Coulomb integrals.
The xc energies, potentials, and kernels for different starting gKS DFT mean-fields are obtained from the \textsc{libxc} library.\cite{marques_cpc2012}
The philosophy behind \textsc{molgw} is to prioritize accuracy and ease of development over computational efficiency, and thus \textsc{molgw} is suitable for small molecular systems.
So \textsc{molgw} solves the random-phase approximation equation [i.e. Eqs.~(\ref{eq:bse_elementsa}-\ref{eq:bse_elementsb}) without the last term]
for the spectral representation of $W$, and thus computes the $GW$ self-energy analytically.
In contrast with other implementations,\cite{blase_prb2011,ren_njp2012} we do not employ auxiliary basis functions to expand the 4-center Coulomb integrals, and so the final $GW$ quasiparticle and BSE excitation energies are exact within the selected basis set.

\begin{figure}[t]
\includegraphics[width=\columnwidth]{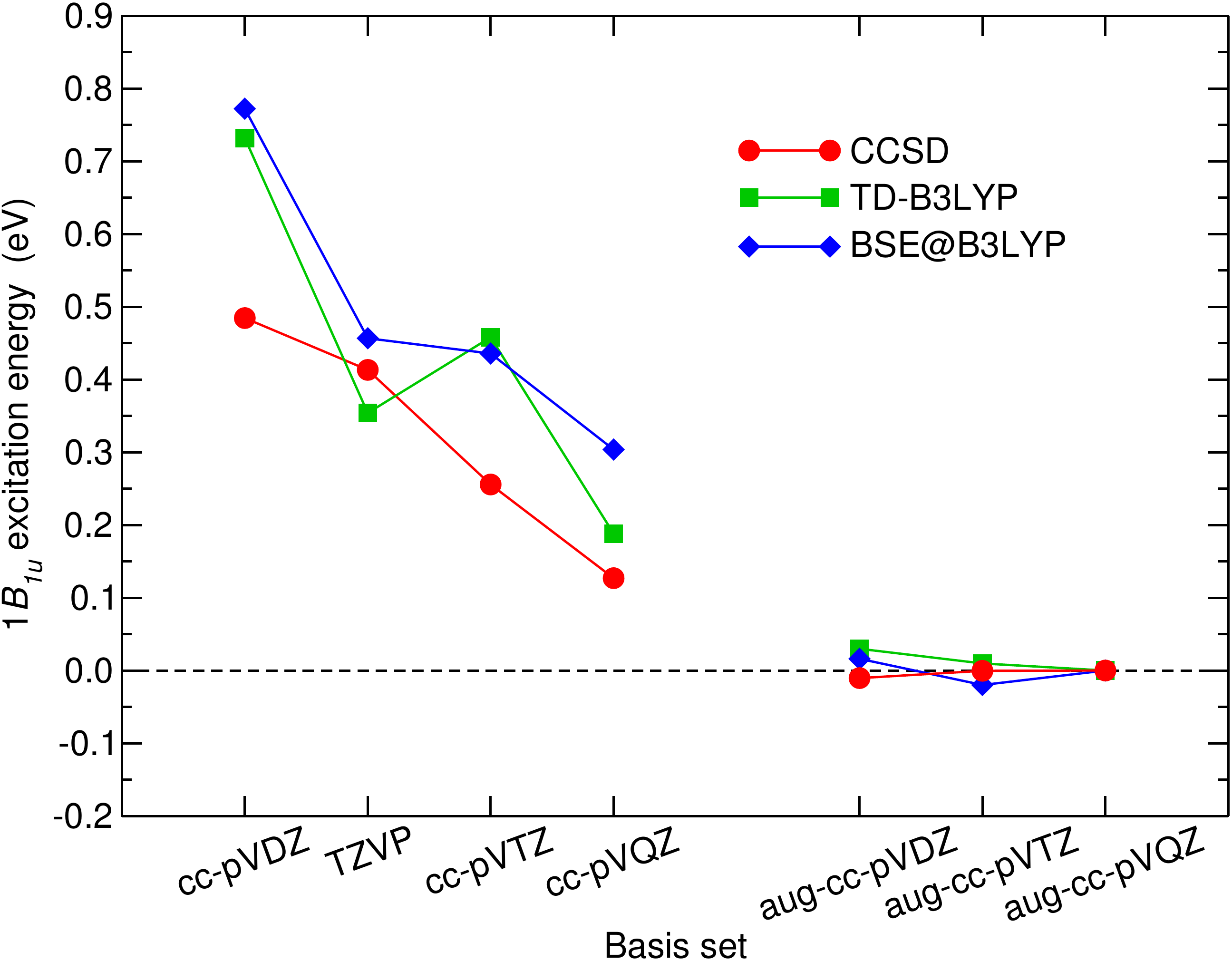}
\caption{Basis set convergence of the first $^1B_{1u}$ excitation in ethene within CCSD (from Ref.~\onlinecite{schreiber_jcp2008}),
from TD-B3LYP, and from BSE based on B3LYP inputs.
The excitation energy for the largest basis set (aug-cc-pVQZ) has been used as the zero for each theoretical approach
\label{fig:conv_ethene_B1u_zero}
}
\end{figure}

Thiel's set contains 28 organic molecules that consist of just four different elements (C, N, O, and H) with the largest molecule being naphthalene C$_{10}$H$_8$. 
The geometries of the molecules in Thiel's set were relaxed within MP2; coordinates for all molecules can be found in the supporting information of Ref.~\onlinecite{schreiber_jcp2008}.
The set comprises tabulated reference excitation energies for 103 singlet and 63 triplet final states. 
The reference data have been obtained from several flavors of coupled-cluster theory, namely CC2, CCSD, and CC3,\cite{koch_jcp1990}, and from complete-active-space second-order perturbation theory (CASPT2).\cite{andersson_jpc1990} The reference excitations of Thiel and coworkers are referred to as ``Best Theoretical Estimates" (BTEs for short), and go beyond a weighted average of the different theoretical approaches. BTEs are theoretical values that have been corrected with some human intuition about the usual discrepancy between these methods and reality. Indeed, BTE values most often lie outside the range of the calculated values. Note that, consistent with the Thiel group's subsequent TD-DFT study,\cite{silvajunior_jcp2008} we disregarded the tabulated double excitation of tetrazine (C$_2$N$_4$H$_2$)
which so far could not be captured by TD-DFT or BSE. This explains why we refer to 103 tabulated values instead of the 104 that appear in the original work.

The original calculations performed on Thiel's set used the so-called TZVP basis set of Alrichs and coworkers.\cite{schafer_jcp1992}
This relatively limited basis was used so that the highly demanding calculations required to build the BTEs were feasible.
For the sake of comparison, we employ the same basis set in our calculations in this work.
The TZVP basis contains 3 series' of valence basis functions, but only one series of polarization functions ($d$ orbitals for C, N, O, and $p$ orbitals for H); it contains no diffuse functions. This basis set yields unconverged results as exemplified in Fig.~\ref{fig:conv_ethene_B1u_zero} for the first excitation in ethene C$_2$H$_4$, which is one of the smallest molecules of the set.
Although all methods considered here (coupled-cluster, TD-DFT, and BSE) are clearly unconverged for the TZVP basis set compared to, for example, the Dunning aug-cc-VQZ basis set,\cite{kendall_jcp1992} it is demonstrated in Fig.~\ref{fig:conv_ethene_B1u_zero} that the convergence rate is similar for the different approaches, justifying the use of a smaller basis.
The deviation of the TZVP value from the converged value ranges from 0.35 to 0.45~eV across the different theoretical methods.
Because the error of all these methods with the TZVP basis agree within 0.1~eV, we expect our calculations to trend across theoretical schemes.
The Thiel group has also shown that the conclusions drawn from the smaller TZVP basis remain valid with the larger aug-cc-pVTZ basis set that includes diffuse functions.
\cite{silvajunior_jcp2010,laurent_ijqc2013}

With these preliminaries done, we are now ready to analyze the performance of BSE for the 28 selected molecules of Thiel's set.

\section{Performance of the BSE for Thiel's set}
\label{secIV}

As shown in Fig.~\ref{fig:bse_flowchart}, the BSE excitation energies rely on eigenenergies and wavefunctions from a prior self-consistent
gKS DFT calculation. As mentioned, a strong dependence of the $GW$ quasiparticle energies on the DFT starting point has previously been discussed in the literature; 
\cite{rostgaard_prb2010,blase_prb2011,marom_prb2012,korzdorfer_prb2012,caruso_prb2012,bruneval_jctc2013}
thus it is not surprising that BSE excitation energies, which are built upon $GW$ quasiparticle energies
(as shown in Fig.~\ref{fig:bse_flowchart}) will also exhibit such a dependence. 
Although the influence of the DFT starting point was mentioned in earlier works,\cite{faber_jcp2013}, to date, no 
systematic quantitative study has been performed.

Hereafter, we will assess the BSE via evaluation of their deviation from the reference BTEs of Thiel's set 
for both singlet and triplet excitations. The BSE is solved using $GW$ quasiparticle energies that have been obtained from several xc approximations to the gKS DFT starting point.
We have selected 4 different xc approximations that are reasonably representative of the popular choices for molecules.
PBE\cite{perdew_prl1996} is a pure semi-local functional with no exact exchange.
B3LYP\cite{stephens_jpc1994} is a hybrid functional containing 20~\%  exact exchange, whose 3 parameters have been adjusted to yield good thermodynamic data, and to this day, is one of the most widely-used functionals in the quantum chemistry community.
BHLYP\cite{becke_jcp1993} is another hybrid functional due to Becke, but contains a significantly larger content of exact exchange, 50~\%. This functional was identified as one of the best starting points for $GW$ in our previous study.\cite{bruneval_jctc2013} Tuned CAM-B3LYP,\cite{okuno_jppa2012} that we label tCAM-B3LYP in the following, is a range-separated hybrid that has the correct full long-range exchange ($\alpha+\beta=1$). It is constructed to yield accurate results in TD-DFT.

The BSE results will then be labeled BSE@PBE, BSE@B3LYP, BSE@BHLYP, and BSE@tCAM-B3LYP respectively.
We reiterate that even though it is not explicitly stated in the short-hand notation, an intermediate $GW$ calculation is always performed.

\subsection{Singlet excitations}

\begin{figure}[t]
\includegraphics[width=\columnwidth]{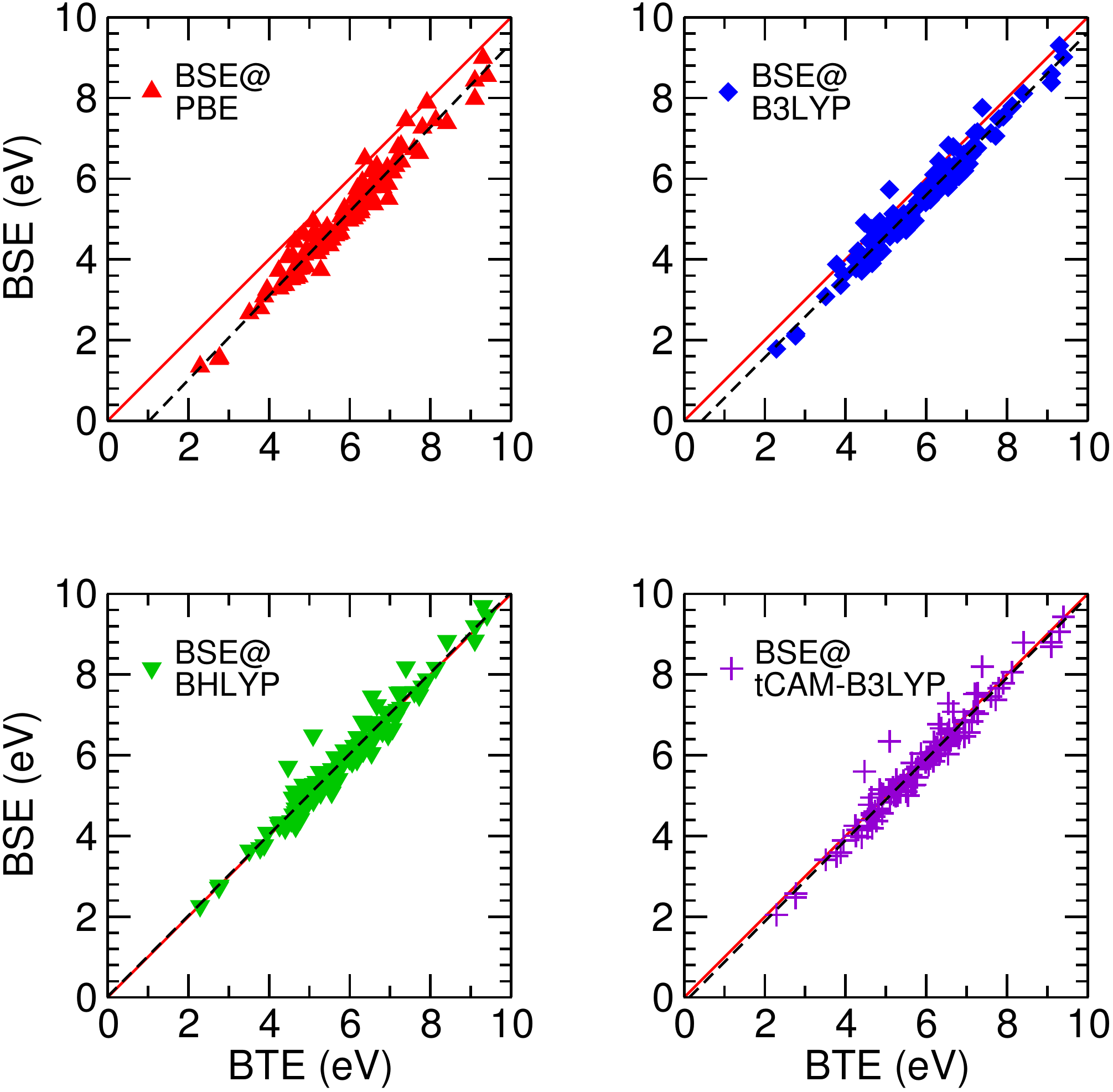}
\caption{Correlation plots for singlet excitations between BSE using different starting points and BTE.
A linear fit of the data is shown with a dashed line.
\label{fig:thiel_singlet_bse}
}
\end{figure}

\begin{figure}[t]
\includegraphics[width=\columnwidth]{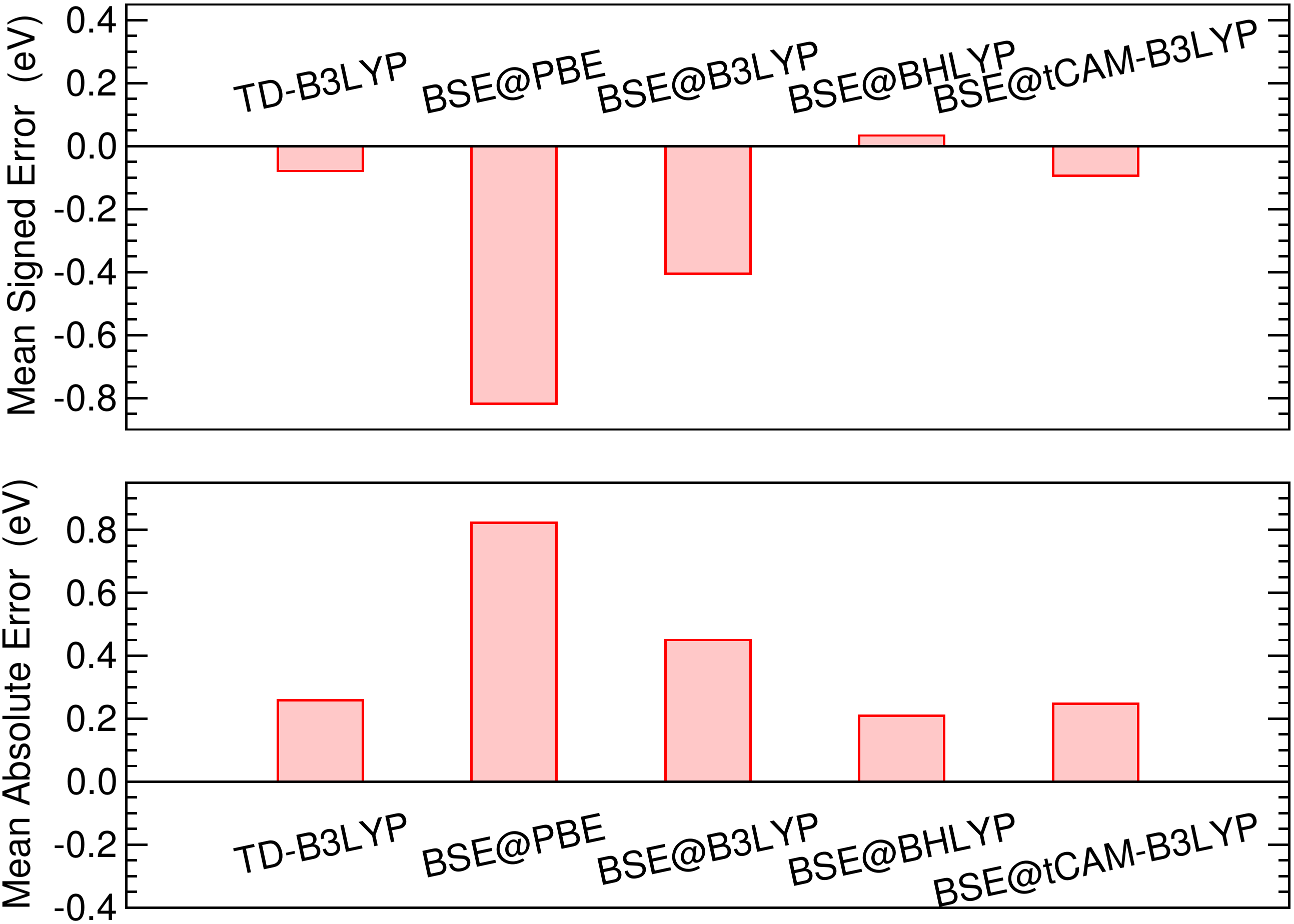}
\caption{Singlet excitations mean signed error (upper panel) and mean absolute error (lower panel) for different schemes.
TD-DFT based on B3LYP (from Ref.~\onlinecite{silvajunior_jcp2008}) is given as a comparison.
\label{fig:error_singlet}
}
\end{figure}

Figure~\ref{fig:thiel_singlet_bse} illustrates the correlation between our computed BSE singlet excitation energies and the reference BTEs evaluated
by Thiel's group.\cite{schreiber_jcp2008}
Perfect agreement would be the case if all points were to lie along the diagonal line.
BSE@PBE consistently yields singlet excitation energies that are too low: almost all data points are below the diagonal.
BSE@B3LYP is much improved but still somewhat underestimates the excitation energies for this set.
BSE@BHLYP and BSE@tCAM-BL3YP, however, are in excellent agreement, with narrow scattering around the diagonal. 
The data in Fig.~\ref{fig:thiel_singlet_bse} remarkably follow the fit by a straight line, whose slope is very close to unity.
This means that for a given starting point the error is quite constant irrespective to the excitation energy.

Thus, whereas semi-local functionals like PBE are not suitable as a starting point for this set of small organic molecules, hybrid functionals do much better, 
and it appears that a larger content of exact exchange improves the agreement with respect to the best theoretical estimates.

In Fig.~\ref{fig:error_singlet}, we report the mean signed error (MSE) and mean absolute error (MAE) with respect to BTEs
for the different approaches considered in this paper, and cite the TD-B3LYP error\cite{silvajunior_jcp2008} as a reference.
We select TD-B3LYP because it performs best for Thiel's set among all TD-DFT xc functionals.\cite{laurent_ijqc2013} In fact, for the type of excitations considered in Thiel's set -- no charge transfer or Rydberg excitations -- TD-B3LYP performs so admirably that we could not expect BSE to outperform it.
As previously noticed, the results reported in Fig.~\ref{fig:error_singlet} show a strong dependence of the BSE excitation error
on the starting point.
More precisely, BSE@PBE underestimates all the excitation energies by almost 1~eV.
BSE@B3LYP also yields excitation energies that are too low. However, with a BHLYP or tCAM-BL3YP starting point, the BSE results can indeed challenge the best TD-DFT excitation energies, yielding results with an MAE of around 0.25~eV.

In conclusion, for singlet excitations, BSE with a properly chosen starting point can be a predictive tool for simple neutral excitations of small organic molecules. We will return to the starting point dependence further in Sec.~\ref{secV}.

\subsection{Triplet excitations}

\begin{figure}[t]
\includegraphics[width=\columnwidth]{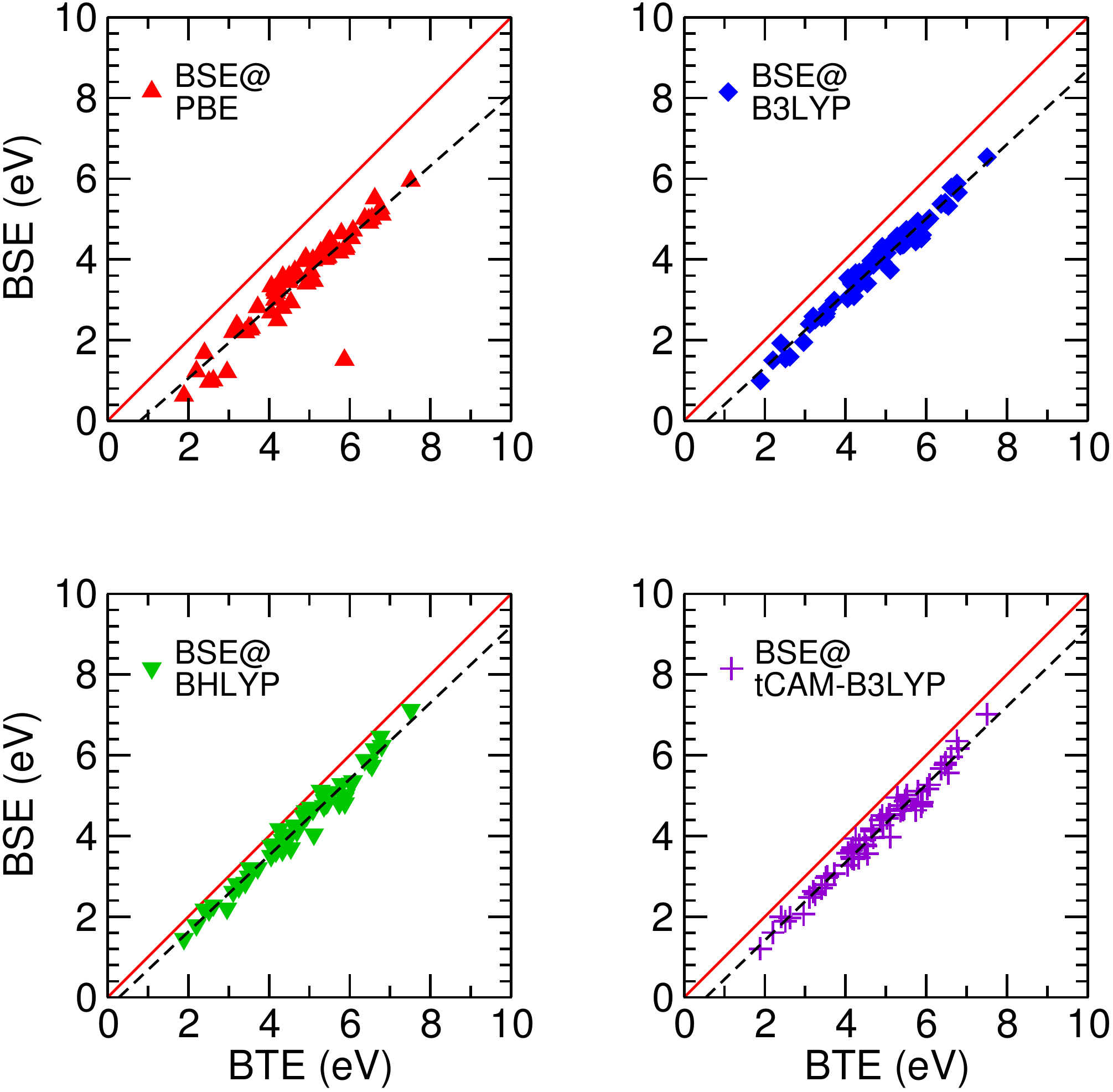}
\caption{Correlation plots for triplet excitations between BSE using different starting points and BTE.
A linear fit of the data is shown with a dashed line.
\label{fig:thiel_triplet_bse}
}
\end{figure}

\begin{figure}[t]
\includegraphics[width=\columnwidth]{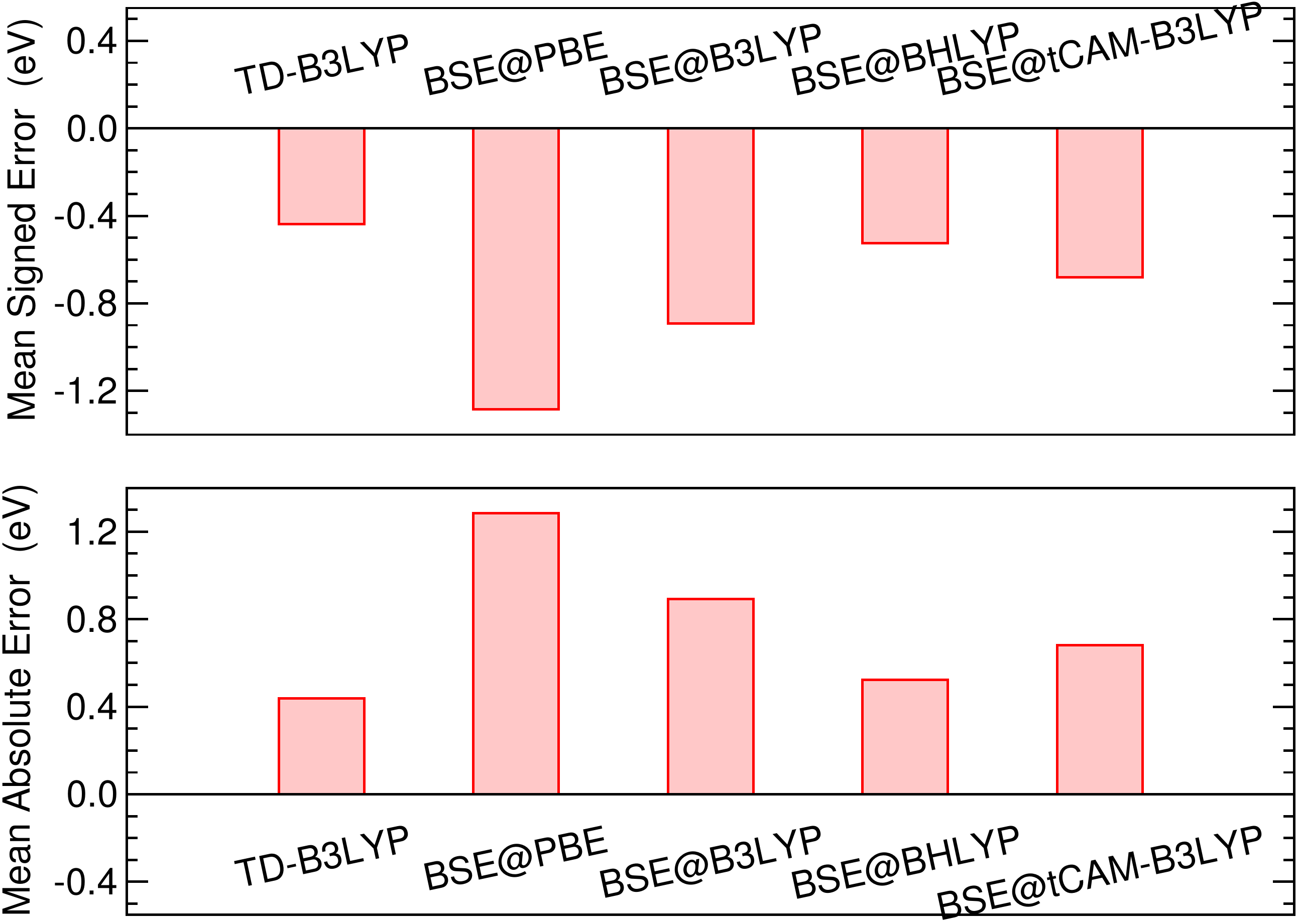}
\caption{Triplet excitations mean signed error (upper panel) and mean absolute error (lower panel) for different schemes.
TD-DFT based on B3LYP (from Ref.~\onlinecite{silvajunior_jcp2008}) is given as a comparison.
\label{fig:error_triplet}
}
\end{figure}

Thiel's set also contains 63 triplet excitation energies, and we now briefly discuss this case.
It is well documented that TD-DFT can have trouble with triplet excitations:\cite{laurent_ijqc2013}
no xc functional of TD-DFT has been able to predict triplet energies of the molecules in Thiel's set at the level obtained for singlets.

Unfortunately, our BSE calculations show a trend very analogous to TD-DFT for triplets.
From the correlation plots shown in Fig.~\ref{fig:thiel_triplet_bse}, we see that
all BSE triplet excitations are too low, regardless of the initial gKS starting point.
Once again, the excitation energies are well fitted by a straight line, however with a slope that departs from unity.
The slope ranging from 0.88 for PBE to 0.97 for tCAM-B3LYP shows that the error is not perfectly constant across the excitation energies:
the larger excitation energies have a greater error.
As expected, BSE@PBE produces the poorest triplet excitation energies of all. Hybrid functionals with some exact exchange (BHLYP and tCAM-BL3YP) improve results relative to the BTE, but the quality of the calculated triplet excitation energies is poorer than for singlets.

The errors shown in Fig.~\ref{fig:error_triplet} confirm that with the best starting point (BHLYP), our BSE calculations match TD-B3LYP in quality, but do not do better for Thiel's set. For both TD-B3LYP and BSE@BHLYP, the error is systematic, with an underestimation of the triplet energies by 0.4~eV. 

\section{Discussion}
\label{secV}

\subsection{A strong dependence on the starting point?}

\begin{figure}[t]
\includegraphics[width=\columnwidth]{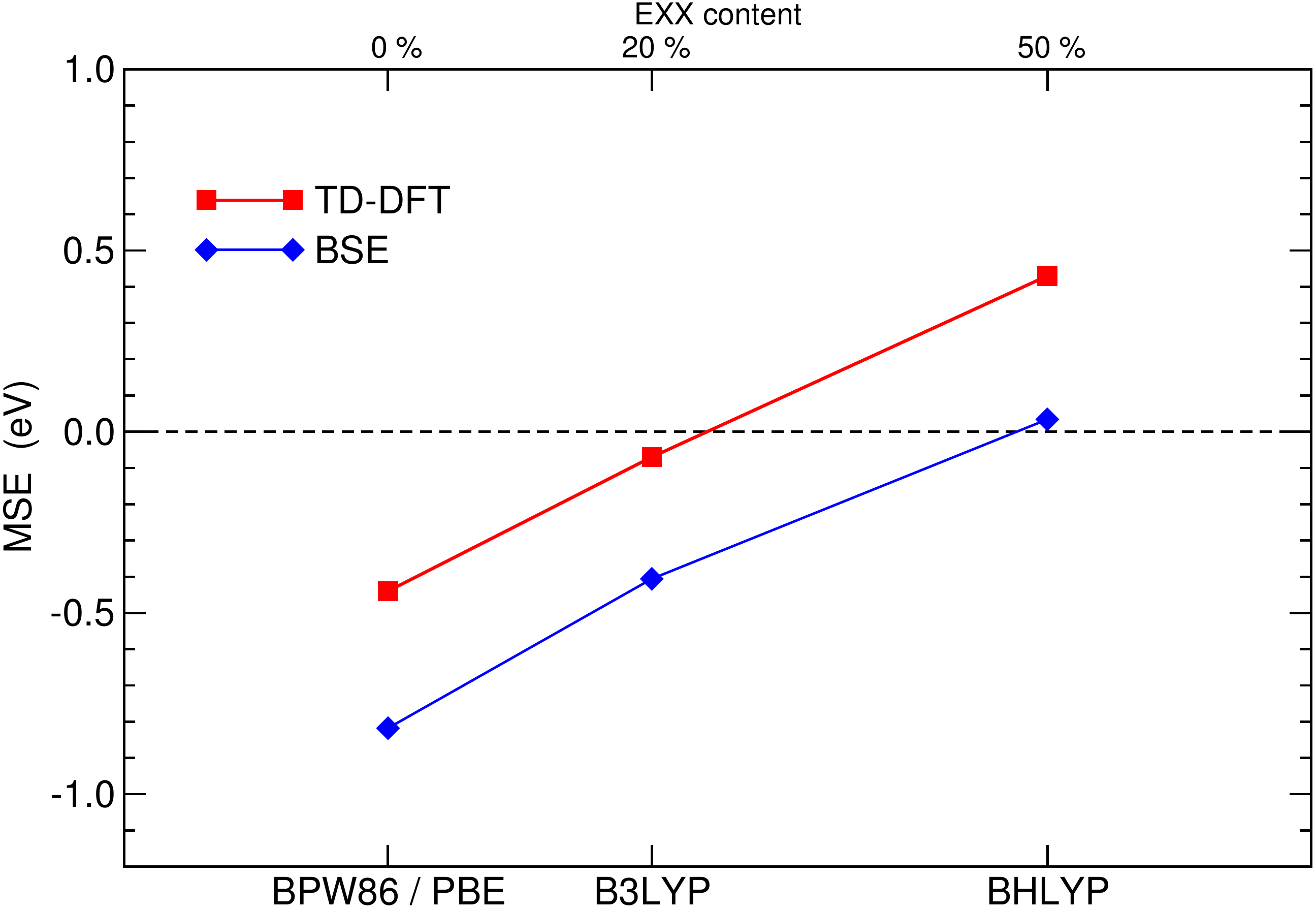}
\caption{Dependence of the mean signed error for the singlet excitation energies of the Thiel's set 
with respect to the content of exact exchange of the underlying xc functional.
The xc functional is used for TD-DFT (red square symbols) or as the starting point of $GW$ and BSE (blue diamond symbols).
\label{fig:DFTxc}
}
\end{figure}

As shown above, the quality of the BSE excitation energies is strongly affected by the gKS starting point. Here, we would like to discuss the sensitivity of the final BSE result to starting point, relative to TD-DFT.

In Fig.~\ref{fig:DFTxc} we represent the mean signed error for Thiel's set as a function of the amount of exact exchange in the xc functional.
The TD-DFT results are from Ref.~\onlinecite{silvajunior_jcp2008}, whereas the BSE results are those reported above.
Both approaches show a noticeable dependence on the exact exchange content, and in the end, they are nearly equally sensitive to the xc functional.

Interestingly, the primary difference between the TD-DFT and BSE schemes lies in the amount of exact exchange that minimizes the error.
TD-DFT performs best with 20-25~\%  exact exchange as in B3LYP or PBE0.\cite{laurent_ijqc2013}
On the contrary, for BSE, the best starting points contain much more exact exchange (around 50~\%).
These two very different optima can be rationalized when decomposing the origin of the errors
in each of the two schemes as we discuss hereafter.

\subsection{Analysis of the origin of errors}

\begin{figure}[t]
\includegraphics[width=\columnwidth]{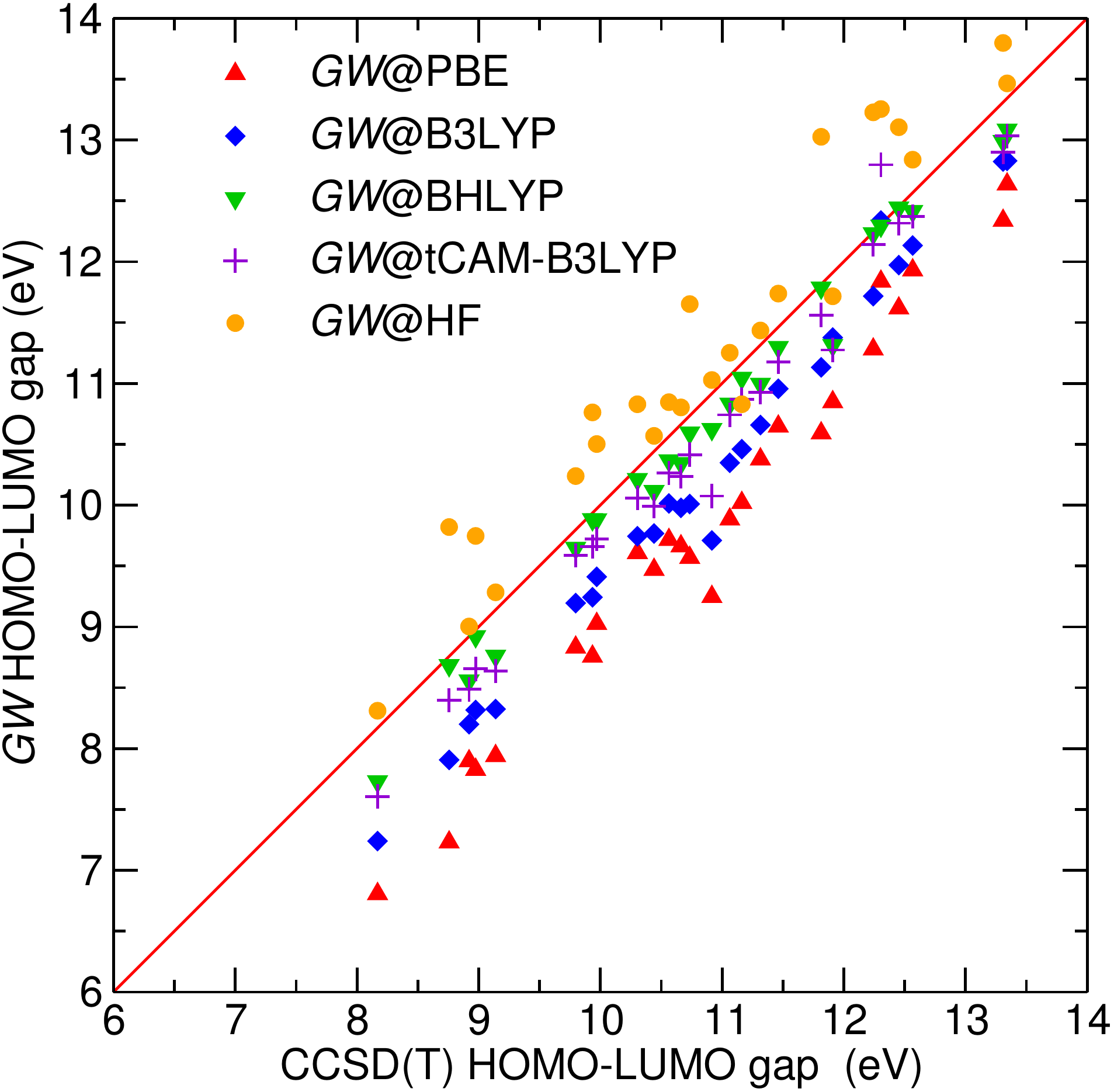}
\caption{Correlation plot for HOMO-LUMO gap from $GW$ and from CCSD(T).
\label{fig:thiel_homolumo_gw}
}
\end{figure}

\begin{figure}[t]
\includegraphics[width=\columnwidth]{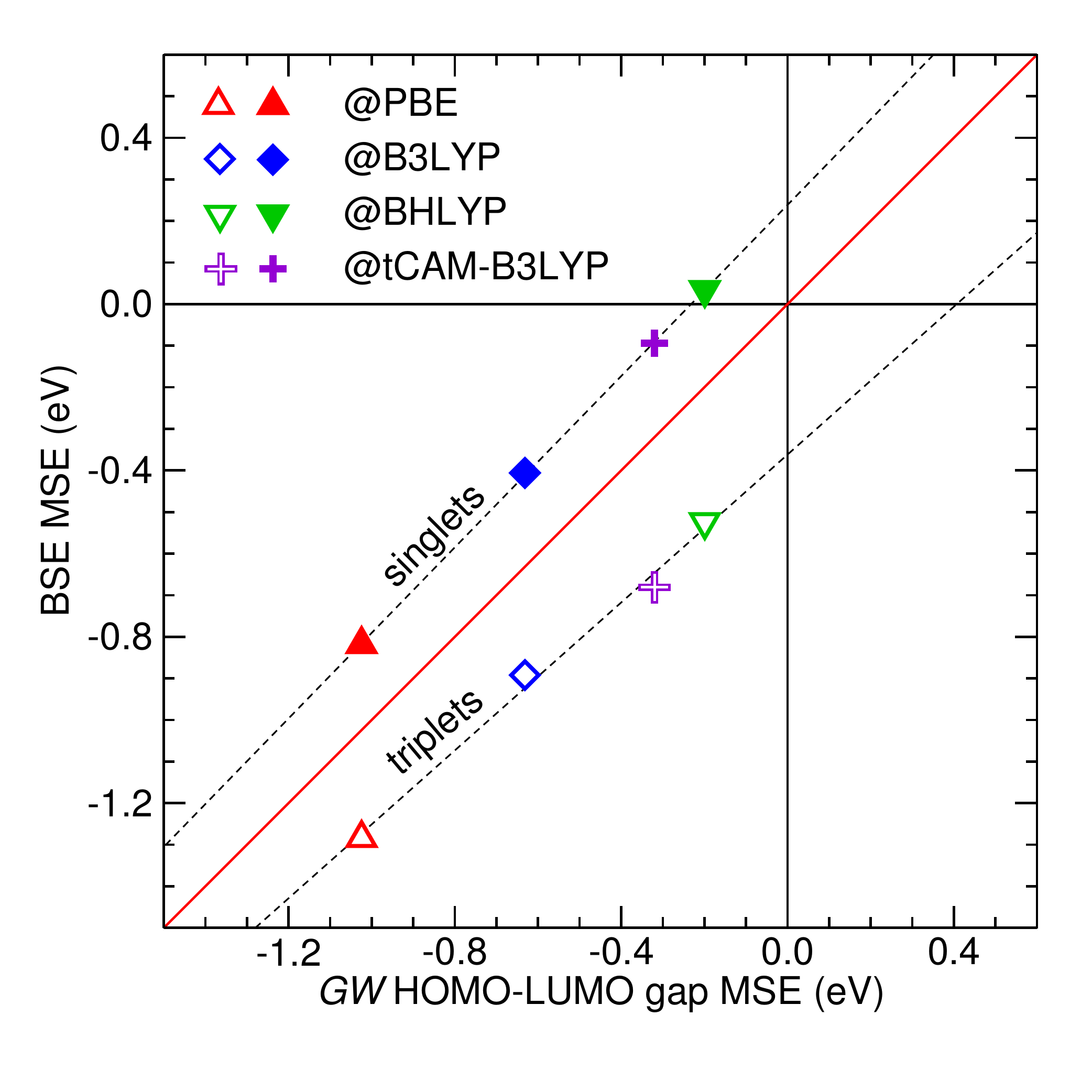}
\caption{Correlation plot between the mean signed error in the $GW$ HOMO-LUMO gap and in the BSE energies
for different gKS starting points.
Both singlet (full symbols) and triplet (open symbols) excitations are represented.
Dashed lines are fits of the singlet or triplet results.
Horizontal and vertical lines mark the zero error lines.
\label{fig:thiel_gw_bse}
}
\end{figure}

As summarized in Fig.~\ref{fig:bse_flowchart}, the BSE energies are obtained from a series of three calculations:
a self-consistent gKS DFT calculation, a $GW$ calculation of the quasiparticle energies, and finally an evaluation of the BSE.
It is interesting to identify which step introduces the noticeable starting point dependence we highlighted above.
Or, another way of posing the same question would be: which among (a) $GW$ quasiparticle energies and (b) the BSE solution is the most sensitive to the gKS input?

To address this question, we need ``Best Theoretical Estimates'' for the quasiparticle energies too.
Although it is not possible to easily access all quasiparticle energies, the HOMO and LUMO energies 
can be obtained via total energy differences with the so-called $\Delta$SCF procedure.
To obtain accurate results, we applied the $\Delta$SCF procedure within CCSD(T), the well-known and standard coupled-cluster method including single and double excitations with triples introduced perturbatively.\cite{bartlett_rmp2007}
All of our coupled-cluster calculations are done with the Gaussian09 code.\cite{gaussian09} We employ the same TZVP basis set used by Thiel, and by us so far in this study. Again, although the diffuse-orbital-less TZVP basis set is, strictly speaking, inadequate for the LUMO wavefunctions for many of these molecules, we use it for consistency with the rest of the calculations performed in this work. The $\Delta$SCF procedure requires three separate total-energy calculations for evaluation of the HOMO-LUMO gap: one for the neutral molecule, and additional calculations for the cation and anion. Note that the underlying Hartree-Fock self-consistent field calculations have been carefully checked against \textsc{molgw}, since the cation and anion cases can be challenging and quite often converge to local minima.

The comparison between $GW$ HOMO-LUMO gaps and CCSD(T) gaps is summarized in Fig.~\ref{fig:thiel_homolumo_gw}.
The results are in line with a previous study of the ionization potentials of small molecules:\cite{bruneval_jctc2013}
$GW$ on PBE largely underestimates the HOMO-LUMO gap, whereas hybrid functionals with a large fraction of exact exchange
do a much better job. Even $GW$ on Hartree-Fock ($GW$@HF) does a decent job for the gaps,
as highlighted in Ref.~\onlinecite{blase_prb2011}. It is worth noting that the trends for the HOMO-LUMO gaps are the same as those for singlet excitations.

Let us quantify this statement by plotting the correlation between the $GW$ MSE and the BSE MSE for different starting points
in Fig.~\ref{fig:thiel_gw_bse}. Our results from BHLYP and tCAM-B3LYP are the most adequate for both singlet and triplet quasiparticle energies: these are the closest to the origin of Fig.~\ref{fig:thiel_gw_bse}. Furthermore, the correlation between the $GW$ and singlet BSE errors is almost perfect: the slope of a line fit to these data is 1.02.
This means that the starting point dependence of BSE is inherited entirely from the $GW$ quasiparticle energies, and for Thiel's set, the BSE singlet excitations are seen to be constant shifts applied to the $GW$ energies. It is unexpected that the details in BSE are so insensitive to the starting point.
Indeed, the screened Coulomb interaction $W$ used in Eqs.~(\ref{eq:bse_elementsa}),(\ref{eq:bse_elementsb}) is
obtained within the random-phase approximation of the underlying gKS DFT calculation, and therefore also varies with starting point. Moreover, $W$ from PBE would be expected to lead to more significant screening than a $W$ constructed from BHLYP, since the HOMO-LUMO gaps of these two approximations
largely differ. But obviously, these differences are too subtle at the short ranges at play in these small molecules to influence the BSE results.
The BSE triplet results can be seen as slightly more correlated to the type of $W$,
but the slope of a line through these data ($\sim 0.9$) is still very close to unity.

Finally, from Fig.~\ref{fig:thiel_gw_bse}, we observe that the intercept of the line fit to the errors differs from zero for both singlets,
0.24~eV, and triplets, -0.36~eV. From this analysis, we can quantify the magnitude of intrinsic errors to BSE.
Indeed, even with a perfect $GW$ approach that produces zero error compared to the CCSD(T) HOMO-LUMO gaps, the BSE singlet/triplet energies would still deviate from those of the best theoretical estimate by 0.2-0.3~eV. 
In addition, the singlets and triplets have opposite signs, indicating that the BSE singlet-triplet splitting can be expected to be systematically overestimated by about 0.6~eV. Furthermore, this conclusion holds independent of starting point since the error lines are nearly parallel.

It is enlightening to carry out the same analysis for TD-DFT.
It is legitimate to do so because in a gKS scheme, be it exact or approximate, the non-local exchange-correlation operator does not have
a derivative discontinuity.\cite{cohen_prb2008,yang_jcp2012}
When combining this statement with the piece-wise linearity of the total energy,  both the gKS HOMO eigenvalue should be equal to minus the IP and
the gKS LUMO eigenvalue should be equal to minus the EA within the exact gKS scheme.
There is no fundamental difference between the IP and the EA.
As a consequence, an accurate xc functional in the gKS scheme should yield frontier eigenvalues that compare well with the BTEs for the HOMO-LUMO gap.

The TD-B3LYP results yield almost perfect singlet excitation energies with a MSE value of -0.08~eV. However, B3LYP produces HOMO and LUMO gKS eigenvalues that strongly deviate from the $\Delta$SCF CCSD(T) reference. The MSE for B3LYP HOMO-LUMO gap is as large as -5.1~eV for the molecules in Thiel's set. Given our BSE results, this strongly suggests that its accurate singlet excitation energies can be ascribed to a significant cancellation of errors. 
Though B3LYP and other similar xc functionals provide a good estimate of singlet excitation energies,
this agreement is not supported by a satisfactory theoretical basis.
Having an xc approximation that yields both correct HOMO-LUMO gaps together with high quality neutral excitations is quite possible, as shown by the
promising recent advances associated with the OT-RSH functional.\cite{refaely_prb2011}

In sum, the dependence of BSE excitation energies on the DFT starting point can be primarily ascribed to the underlying $GW$ quasiparticle energies.
The BSE step itself has a systematic bias towards an overestimation of the singlet energies and an underestimation of the triplet energies. Although both errors are rather small (0.2-0.4~eV), they lead to overestimated singlet-triplet splittings. However, the BSE error in the singlet-triplet splitting is of the same order of magnitude as the best TD-DFT schemes.
Finally, the agreement of the TD-DFT approximation with the reference singlets has been shown to rely on a cancellation of significant errors, a cancellation that may be less complete by excitations that deviate from the simple nature of those considered here, for example in molecules more complex than for the small molecules in Thiel's set.

\section{Conclusion}

In this article, we have evaluated the performance of BSE for singlet and triplet excitation energies
of Thiel's set of 28 small organic molecules.
The quality of our BSE results is found to be sensitive to the chosen DFT gKS starting point.
Semi-local starting points, such as PBE, fail significantly for the molecules in Thiel's set.
Hybrid functional starting points in general produce much better results.
Among the hybrid functionals, those containing a large contribution from exact exchange (BHLYP or tCAM-BL3YP)
perform best. We would advocate the use of such functionals (or similar ones) in future BSE studies.

The dependence on the starting point can be connected to the different content of exact exchange in each functional.
This sensitivity, though important, is not more significant than the one observed for xc functionals in TD-DFT.
The performance of BSE for singlets is clearly superior to its performance for triplets.
The same statement holds for TD-DFT.

When analyzing the origin of the error with HOMO-LUMO gaps evaluated from $\Delta$SCF based on CCSD(T),
we found that the entire BSE starting point dependence originates with the $GW$ quasiparticle energies.
The details of screened Coulomb interaction $W$ used in the BSE kernel is not significant for the small molecules in Thiel's set.
Thus, maximizing the accuracy in quasiparticle energies should minimize the error in BSE. However, residual errors in BSE are non-vanishing. For the best quasiparticle energies, we predict that singlets would be over-estimated by 0.2~eV and triplets would be underestimated by $-0.4$~eV.
Future investigation of the limitations of the BSE for singlet-triplet splitting would be desirable.

\begin{acknowledgments}
F.B. acknowledges the Enhanced Eurotalent program and the France Berkeley Fund for supporting his sabbatical leave in UC Berkeley.
This work is supported by the U.S. Department of Energy, Office of Basic Energy Sciences and of Advanced Scientific Computing Research through the SciDAC Program on Excited State Phenomena.
S.M.H. is supported by the Chemical Sciences, Geosciences, and Biosciences Division in Office of Basic Energy Sciences of the U.S. Department of Energy.
Portions of this work took place at the Molecular Foundry, supported by the U.S. Department of Energy, Office of Basic Energy Sciences.
This work was performed using HPC resources from GENCI-CCRT-TGCC (Grants No. 2014-096018). 
\end{acknowledgments}


\begin{thebibliography}{68}%
\makeatletter
\providecommand \@ifxundefined [1]{%
 \@ifx{#1\undefined}
}%
\providecommand \@ifnum [1]{%
 \ifnum #1\expandafter \@firstoftwo
 \else \expandafter \@secondoftwo
 \fi
}%
\providecommand \@ifx [1]{%
 \ifx #1\expandafter \@firstoftwo
 \else \expandafter \@secondoftwo
 \fi
}%
\providecommand \natexlab [1]{#1}%
\providecommand \enquote  [1]{``#1''}%
\providecommand \bibnamefont  [1]{#1}%
\providecommand \bibfnamefont [1]{#1}%
\providecommand \citenamefont [1]{#1}%
\providecommand \href@noop [0]{\@secondoftwo}%
\providecommand \href [0]{\begingroup \@sanitize@url \@href}%
\providecommand \@href[1]{\@@startlink{#1}\@@href}%
\providecommand \@@href[1]{\endgroup#1\@@endlink}%
\providecommand \@sanitize@url [0]{\catcode `\\12\catcode `\$12\catcode
  `\&12\catcode `\#12\catcode `\^12\catcode `\_12\catcode `\%12\relax}%
\providecommand \@@startlink[1]{}%
\providecommand \@@endlink[0]{}%
\providecommand \url  [0]{\begingroup\@sanitize@url \@url }%
\providecommand \@url [1]{\endgroup\@href {#1}{\urlprefix }}%
\providecommand \urlprefix  [0]{URL }%
\providecommand \Eprint [0]{\href }%
\providecommand \doibase [0]{http://dx.doi.org/}%
\providecommand \selectlanguage [0]{\@gobble}%
\providecommand \bibinfo  [0]{\@secondoftwo}%
\providecommand \bibfield  [0]{\@secondoftwo}%
\providecommand \translation [1]{[#1]}%
\providecommand \BibitemOpen [0]{}%
\providecommand \bibitemStop [0]{}%
\providecommand \bibitemNoStop [0]{.\EOS\space}%
\providecommand \EOS [0]{\spacefactor3000\relax}%
\providecommand \BibitemShut  [1]{\csname bibitem#1\endcsname}%
\let\auto@bib@innerbib\@empty
\bibitem [{\citenamefont {Runge}\ and\ \citenamefont
  {Gross}(1984)}]{runge_prl1984}%
  \BibitemOpen
  \bibfield  {author} {\bibinfo {author} {\bibfnamefont {E.}~\bibnamefont
  {Runge}}\ and\ \bibinfo {author} {\bibfnamefont {E.~K.~U.}\ \bibnamefont
  {Gross}},\ }\href {\doibase 10.1103/PhysRevLett.52.997} {\bibfield  {journal}
  {\bibinfo  {journal} {Phys. Rev. Lett.}\ }\textbf {\bibinfo {volume} {52}},\
  \bibinfo {pages} {997} (\bibinfo {year} {1984})}\BibitemShut {NoStop}%
\bibitem [{\citenamefont {Marques}\ and\ \citenamefont
  {Gross}(2004)}]{marques_arpc2004}%
  \BibitemOpen
  \bibfield  {author} {\bibinfo {author} {\bibfnamefont {M.}~\bibnamefont
  {Marques}}\ and\ \bibinfo {author} {\bibfnamefont {E.}~\bibnamefont
  {Gross}},\ }\href {\doibase 10.1146/annurev.physchem.55.091602.094449}
  {\bibfield  {journal} {\bibinfo  {journal} {Annu. Rev. Phys. Chem.}\ }\textbf
  {\bibinfo {volume} {55}},\ \bibinfo {pages} {427} (\bibinfo {year} {2004})},\
  \bibinfo {note} {pMID: 15117259},\ \Eprint
  {http://arxiv.org/abs/http://dx.doi.org/10.1146/annurev.physchem.55.091602.094449}
  {http://dx.doi.org/10.1146/annurev.physchem.55.091602.094449} \BibitemShut
  {NoStop}%
\bibitem [{\citenamefont {Strinati}(1988)}]{strinati_rnc1988}%
  \BibitemOpen
  \bibfield  {author} {\bibinfo {author} {\bibfnamefont {G.}~\bibnamefont
  {Strinati}},\ }\href {\doibase 10.1007/BF02725962} {\bibfield  {journal}
  {\bibinfo  {journal} {Riv. Nuovo Cimento}\ }\textbf {\bibinfo {volume}
  {11}},\ \bibinfo {pages} {1} (\bibinfo {year} {1988})}\BibitemShut {NoStop}%
\bibitem [{\citenamefont {Onida}, \citenamefont {Reining},\ and\ \citenamefont
  {Rubio}(2002)}]{onida_rmp2002}%
  \BibitemOpen
  \bibfield  {author} {\bibinfo {author} {\bibfnamefont {G.}~\bibnamefont
  {Onida}}, \bibinfo {author} {\bibfnamefont {L.}~\bibnamefont {Reining}}, \
  and\ \bibinfo {author} {\bibfnamefont {A.}~\bibnamefont {Rubio}},\ }\href
  {\doibase 10.1103/RevModPhys.74.601} {\bibfield  {journal} {\bibinfo
  {journal} {Rev. Mod. Phys.}\ }\textbf {\bibinfo {volume} {74}},\ \bibinfo
  {pages} {601} (\bibinfo {year} {2002})}\BibitemShut {NoStop}%
\bibitem [{\citenamefont {Bauernschmitt}\ and\ \citenamefont
  {Ahlrichs}(1996)}]{bauernschmitt_cpl1996}%
  \BibitemOpen
  \bibfield  {author} {\bibinfo {author} {\bibfnamefont {R.}~\bibnamefont
  {Bauernschmitt}}\ and\ \bibinfo {author} {\bibfnamefont {R.}~\bibnamefont
  {Ahlrichs}},\ }\href {\doibase
  http://dx.doi.org/10.1016/0009-2614(96)00440-X} {\bibfield  {journal}
  {\bibinfo  {journal} {Chem. Phys. Lett.}\ }\textbf {\bibinfo {volume}
  {256}},\ \bibinfo {pages} {454 } (\bibinfo {year} {1996})}\BibitemShut
  {NoStop}%
\bibitem [{\citenamefont {Tozer}\ and\ \citenamefont
  {Handy}(1998)}]{tozer_jcp1998}%
  \BibitemOpen
  \bibfield  {author} {\bibinfo {author} {\bibfnamefont {D.~J.}\ \bibnamefont
  {Tozer}}\ and\ \bibinfo {author} {\bibfnamefont {N.~C.}\ \bibnamefont
  {Handy}},\ }\href {\doibase http://dx.doi.org/10.1063/1.477711} {\bibfield
  {journal} {\bibinfo  {journal} {J. Chem. Phys.}\ }\textbf {\bibinfo {volume}
  {109}},\ \bibinfo {pages} {10180} (\bibinfo {year} {1998})}\BibitemShut
  {NoStop}%
\bibitem [{\citenamefont {Dreuw}\ and\ \citenamefont
  {Head-Gordon}(2004)}]{dreuw_jacs2004}%
  \BibitemOpen
  \bibfield  {author} {\bibinfo {author} {\bibfnamefont {A.}~\bibnamefont
  {Dreuw}}\ and\ \bibinfo {author} {\bibfnamefont {M.}~\bibnamefont
  {Head-Gordon}},\ }\href {\doibase 10.1021/ja039556n} {\bibfield  {journal}
  {\bibinfo  {journal} {J. Am. Chem. Soc.}\ }\textbf {\bibinfo {volume}
  {126}},\ \bibinfo {pages} {4007} (\bibinfo {year} {2004})},\ \bibinfo {note}
  {pMID: 15038755},\ \Eprint
  {http://arxiv.org/abs/http://dx.doi.org/10.1021/ja039556n}
  {http://dx.doi.org/10.1021/ja039556n} \BibitemShut {NoStop}%
\bibitem [{\citenamefont {Okuno}\ \emph {et~al.}(2012)\citenamefont {Okuno},
  \citenamefont {Shigeta}, \citenamefont {Kishi}, \citenamefont {Miyasaka},\
  and\ \citenamefont {Nakano}}]{okuno_jppa2012}%
  \BibitemOpen
  \bibfield  {author} {\bibinfo {author} {\bibfnamefont {K.}~\bibnamefont
  {Okuno}}, \bibinfo {author} {\bibfnamefont {Y.}~\bibnamefont {Shigeta}},
  \bibinfo {author} {\bibfnamefont {R.}~\bibnamefont {Kishi}}, \bibinfo
  {author} {\bibfnamefont {H.}~\bibnamefont {Miyasaka}}, \ and\ \bibinfo
  {author} {\bibfnamefont {M.}~\bibnamefont {Nakano}},\ }\href {\doibase
  http://dx.doi.org/10.1016/j.jphotochem.2012.03.003} {\bibfield  {journal}
  {\bibinfo  {journal} {J. Photochem. Photobiol. A}\ }\textbf {\bibinfo
  {volume} {235}},\ \bibinfo {pages} {29 } (\bibinfo {year}
  {2012})}\BibitemShut {NoStop}%
\bibitem [{\citenamefont {Baer}\ and\ \citenamefont
  {Neuhauser}(2005)}]{baer_prl2005}%
  \BibitemOpen
  \bibfield  {author} {\bibinfo {author} {\bibfnamefont {R.}~\bibnamefont
  {Baer}}\ and\ \bibinfo {author} {\bibfnamefont {D.}~\bibnamefont
  {Neuhauser}},\ }\href {\doibase 10.1103/PhysRevLett.94.043002} {\bibfield
  {journal} {\bibinfo  {journal} {Phys. Rev. Lett.}\ }\textbf {\bibinfo
  {volume} {94}},\ \bibinfo {pages} {043002} (\bibinfo {year}
  {2005})}\BibitemShut {NoStop}%
\bibitem [{\citenamefont {Livshits}\ and\ \citenamefont
  {Baer}(2007)}]{livshits_pccp2007}%
  \BibitemOpen
  \bibfield  {author} {\bibinfo {author} {\bibfnamefont {E.}~\bibnamefont
  {Livshits}}\ and\ \bibinfo {author} {\bibfnamefont {R.}~\bibnamefont
  {Baer}},\ }\href {\doibase 10.1039/B617919C} {\bibfield  {journal} {\bibinfo
  {journal} {Phys. Chem. Chem. Phys.}\ }\textbf {\bibinfo {volume} {9}},\
  \bibinfo {pages} {2932} (\bibinfo {year} {2007})}\BibitemShut {NoStop}%
\bibitem [{\citenamefont {Refaely-Abramson}\ \emph {et~al.}(2012)\citenamefont
  {Refaely-Abramson}, \citenamefont {Sharifzadeh}, \citenamefont {Govind},
  \citenamefont {Autschbach}, \citenamefont {Neaton}, \citenamefont {Baer},\
  and\ \citenamefont {Kronik}}]{refaely_prl2012}%
  \BibitemOpen
  \bibfield  {author} {\bibinfo {author} {\bibfnamefont {S.}~\bibnamefont
  {Refaely-Abramson}}, \bibinfo {author} {\bibfnamefont {S.}~\bibnamefont
  {Sharifzadeh}}, \bibinfo {author} {\bibfnamefont {N.}~\bibnamefont {Govind}},
  \bibinfo {author} {\bibfnamefont {J.}~\bibnamefont {Autschbach}}, \bibinfo
  {author} {\bibfnamefont {J.~B.}\ \bibnamefont {Neaton}}, \bibinfo {author}
  {\bibfnamefont {R.}~\bibnamefont {Baer}}, \ and\ \bibinfo {author}
  {\bibfnamefont {L.}~\bibnamefont {Kronik}},\ }\href {\doibase
  10.1103/PhysRevLett.109.226405} {\bibfield  {journal} {\bibinfo  {journal}
  {Phys. Rev. Lett.}\ }\textbf {\bibinfo {volume} {109}},\ \bibinfo {pages}
  {226405} (\bibinfo {year} {2012})}\BibitemShut {NoStop}%
\bibitem [{\citenamefont {Sottile}, \citenamefont {Olevano},\ and\
  \citenamefont {Reining}(2003)}]{sottile_prl2003}%
  \BibitemOpen
  \bibfield  {author} {\bibinfo {author} {\bibfnamefont {F.}~\bibnamefont
  {Sottile}}, \bibinfo {author} {\bibfnamefont {V.}~\bibnamefont {Olevano}}, \
  and\ \bibinfo {author} {\bibfnamefont {L.}~\bibnamefont {Reining}},\ }\href
  {\doibase 10.1103/PhysRevLett.91.056402} {\bibfield  {journal} {\bibinfo
  {journal} {Phys. Rev. Lett.}\ }\textbf {\bibinfo {volume} {91}},\ \bibinfo
  {pages} {056402} (\bibinfo {year} {2003})}\BibitemShut {NoStop}%
\bibitem [{\citenamefont {Marini}, \citenamefont {Del~Sole},\ and\
  \citenamefont {Rubio}(2003)}]{marini_prl2003}%
  \BibitemOpen
  \bibfield  {author} {\bibinfo {author} {\bibfnamefont {A.}~\bibnamefont
  {Marini}}, \bibinfo {author} {\bibfnamefont {R.}~\bibnamefont {Del~Sole}}, \
  and\ \bibinfo {author} {\bibfnamefont {A.}~\bibnamefont {Rubio}},\ }\href
  {\doibase 10.1103/PhysRevLett.91.256402} {\bibfield  {journal} {\bibinfo
  {journal} {Phys. Rev. Lett.}\ }\textbf {\bibinfo {volume} {91}},\ \bibinfo
  {pages} {256402} (\bibinfo {year} {2003})}\BibitemShut {NoStop}%
\bibitem [{\citenamefont {Sharma}\ \emph {et~al.}(2011)\citenamefont {Sharma},
  \citenamefont {Dewhurst}, \citenamefont {Sanna},\ and\ \citenamefont
  {Gross}}]{sharma_prl2011}%
  \BibitemOpen
  \bibfield  {author} {\bibinfo {author} {\bibfnamefont {S.}~\bibnamefont
  {Sharma}}, \bibinfo {author} {\bibfnamefont {J.~K.}\ \bibnamefont
  {Dewhurst}}, \bibinfo {author} {\bibfnamefont {A.}~\bibnamefont {Sanna}}, \
  and\ \bibinfo {author} {\bibfnamefont {E.~K.~U.}\ \bibnamefont {Gross}},\
  }\href {\doibase 10.1103/PhysRevLett.107.186401} {\bibfield  {journal}
  {\bibinfo  {journal} {Phys. Rev. Lett.}\ }\textbf {\bibinfo {volume} {107}},\
  \bibinfo {pages} {186401} (\bibinfo {year} {2011})}\BibitemShut {NoStop}%
\bibitem [{\citenamefont {Botti}\ \emph {et~al.}(2007)\citenamefont {Botti},
  \citenamefont {Schindlmayr}, \citenamefont {Sole},\ and\ \citenamefont
  {Reining}}]{botti_rpp2007}%
  \BibitemOpen
  \bibfield  {author} {\bibinfo {author} {\bibfnamefont {S.}~\bibnamefont
  {Botti}}, \bibinfo {author} {\bibfnamefont {A.}~\bibnamefont {Schindlmayr}},
  \bibinfo {author} {\bibfnamefont {R.~D.}\ \bibnamefont {Sole}}, \ and\
  \bibinfo {author} {\bibfnamefont {L.}~\bibnamefont {Reining}},\ }\href
  {http://stacks.iop.org/0034-4885/70/i=3/a=R02} {\bibfield  {journal}
  {\bibinfo  {journal} {Rep. Prog. Phys.}\ }\textbf {\bibinfo {volume} {70}},\
  \bibinfo {pages} {357} (\bibinfo {year} {2007})}\BibitemShut {NoStop}%
\bibitem [{\citenamefont {Bruneval}\ \emph {et~al.}(2006)\citenamefont
  {Bruneval}, \citenamefont {Sottile}, \citenamefont {Olevano},\ and\
  \citenamefont {Reining}}]{bruneval_jcp2006}%
  \BibitemOpen
  \bibfield  {author} {\bibinfo {author} {\bibfnamefont {F.}~\bibnamefont
  {Bruneval}}, \bibinfo {author} {\bibfnamefont {F.}~\bibnamefont {Sottile}},
  \bibinfo {author} {\bibfnamefont {V.}~\bibnamefont {Olevano}}, \ and\
  \bibinfo {author} {\bibfnamefont {L.}~\bibnamefont {Reining}},\ }\href
  {\doibase 10.1063/1.2186996} {\bibfield  {journal} {\bibinfo  {journal} {J.
  Chem. Phys.}\ }\textbf {\bibinfo {volume} {124}},\ \bibinfo {pages} {144113}
  (\bibinfo {year} {2006})}\BibitemShut {NoStop}%
\bibitem [{\citenamefont {Hanke}\ and\ \citenamefont
  {Sham}(1975)}]{hanke_prb1975}%
  \BibitemOpen
  \bibfield  {author} {\bibinfo {author} {\bibfnamefont {W.}~\bibnamefont
  {Hanke}}\ and\ \bibinfo {author} {\bibfnamefont {L.~J.}\ \bibnamefont
  {Sham}},\ }\href {\doibase 10.1103/PhysRevB.12.4501} {\bibfield  {journal}
  {\bibinfo  {journal} {Phys. Rev. B}\ }\textbf {\bibinfo {volume} {12}},\
  \bibinfo {pages} {4501} (\bibinfo {year} {1975})}\BibitemShut {NoStop}%
\bibitem [{\citenamefont {Albrecht}\ \emph {et~al.}(1998)\citenamefont
  {Albrecht}, \citenamefont {Reining}, \citenamefont {Del~Sole},\ and\
  \citenamefont {Onida}}]{albrecht_prl1998}%
  \BibitemOpen
  \bibfield  {author} {\bibinfo {author} {\bibfnamefont {S.}~\bibnamefont
  {Albrecht}}, \bibinfo {author} {\bibfnamefont {L.}~\bibnamefont {Reining}},
  \bibinfo {author} {\bibfnamefont {R.}~\bibnamefont {Del~Sole}}, \ and\
  \bibinfo {author} {\bibfnamefont {G.}~\bibnamefont {Onida}},\ }\href
  {\doibase 10.1103/PhysRevLett.80.4510} {\bibfield  {journal} {\bibinfo
  {journal} {Phys. Rev. Lett.}\ }\textbf {\bibinfo {volume} {80}},\ \bibinfo
  {pages} {4510} (\bibinfo {year} {1998})}\BibitemShut {NoStop}%
\bibitem [{\citenamefont {Shirley}(1998)}]{shirley_prl1998}%
  \BibitemOpen
  \bibfield  {author} {\bibinfo {author} {\bibfnamefont {E.~L.}\ \bibnamefont
  {Shirley}},\ }\href {\doibase 10.1103/PhysRevLett.80.794} {\bibfield
  {journal} {\bibinfo  {journal} {Phys. Rev. Lett.}\ }\textbf {\bibinfo
  {volume} {80}},\ \bibinfo {pages} {794} (\bibinfo {year} {1998})}\BibitemShut
  {NoStop}%
\bibitem [{\citenamefont {Rohlfing}\ and\ \citenamefont
  {Louie}(1998)}]{rohlfing_prl1998}%
  \BibitemOpen
  \bibfield  {author} {\bibinfo {author} {\bibfnamefont {M.}~\bibnamefont
  {Rohlfing}}\ and\ \bibinfo {author} {\bibfnamefont {S.~G.}\ \bibnamefont
  {Louie}},\ }\href {\doibase 10.1103/PhysRevLett.81.2312} {\bibfield
  {journal} {\bibinfo  {journal} {Phys. Rev. Lett.}\ }\textbf {\bibinfo
  {volume} {81}},\ \bibinfo {pages} {2312} (\bibinfo {year}
  {1998})}\BibitemShut {NoStop}%
\bibitem [{\citenamefont {Spataru}\ \emph {et~al.}(2004)\citenamefont
  {Spataru}, \citenamefont {Ismail-Beigi}, \citenamefont {Benedict},\ and\
  \citenamefont {Louie}}]{spataru_prl2004}%
  \BibitemOpen
  \bibfield  {author} {\bibinfo {author} {\bibfnamefont {C.~D.}\ \bibnamefont
  {Spataru}}, \bibinfo {author} {\bibfnamefont {S.}~\bibnamefont
  {Ismail-Beigi}}, \bibinfo {author} {\bibfnamefont {L.~X.}\ \bibnamefont
  {Benedict}}, \ and\ \bibinfo {author} {\bibfnamefont {S.~G.}\ \bibnamefont
  {Louie}},\ }\href {\doibase 10.1103/PhysRevLett.92.077402} {\bibfield
  {journal} {\bibinfo  {journal} {Phys. Rev. Lett.}\ }\textbf {\bibinfo
  {volume} {92}},\ \bibinfo {pages} {077402} (\bibinfo {year}
  {2004})}\BibitemShut {NoStop}%
\bibitem [{\citenamefont {Blase}, \citenamefont {Attaccalite},\ and\
  \citenamefont {Olevano}(2011)}]{blase_prb2011}%
  \BibitemOpen
  \bibfield  {author} {\bibinfo {author} {\bibfnamefont {X.}~\bibnamefont
  {Blase}}, \bibinfo {author} {\bibfnamefont {C.}~\bibnamefont {Attaccalite}},
  \ and\ \bibinfo {author} {\bibfnamefont {V.}~\bibnamefont {Olevano}},\ }\href
  {\doibase 10.1103/PhysRevB.83.115103} {\bibfield  {journal} {\bibinfo
  {journal} {Phys. Rev. B}\ }\textbf {\bibinfo {volume} {83}},\ \bibinfo
  {pages} {115103} (\bibinfo {year} {2011})}\BibitemShut {NoStop}%
\bibitem [{\citenamefont {Faber}\ \emph {et~al.}(2011)\citenamefont {Faber},
  \citenamefont {Attaccalite}, \citenamefont {Olevano}, \citenamefont {Runge},\
  and\ \citenamefont {Blase}}]{faber_prb2011}%
  \BibitemOpen
  \bibfield  {author} {\bibinfo {author} {\bibfnamefont {C.}~\bibnamefont
  {Faber}}, \bibinfo {author} {\bibfnamefont {C.}~\bibnamefont {Attaccalite}},
  \bibinfo {author} {\bibfnamefont {V.}~\bibnamefont {Olevano}}, \bibinfo
  {author} {\bibfnamefont {E.}~\bibnamefont {Runge}}, \ and\ \bibinfo {author}
  {\bibfnamefont {X.}~\bibnamefont {Blase}},\ }\href {\doibase
  10.1103/PhysRevB.83.115123} {\bibfield  {journal} {\bibinfo  {journal} {Phys.
  Rev. B}\ }\textbf {\bibinfo {volume} {83}},\ \bibinfo {pages} {115123}
  (\bibinfo {year} {2011})}\BibitemShut {NoStop}%
\bibitem [{\citenamefont {Sharifzadeh}\ \emph
  {et~al.}(2012{\natexlab{a}})\citenamefont {Sharifzadeh}, \citenamefont
  {Biller}, \citenamefont {Kronik},\ and\ \citenamefont
  {Neaton}}]{sharifzadeh_prb2012}%
  \BibitemOpen
  \bibfield  {author} {\bibinfo {author} {\bibfnamefont {S.}~\bibnamefont
  {Sharifzadeh}}, \bibinfo {author} {\bibfnamefont {A.}~\bibnamefont {Biller}},
  \bibinfo {author} {\bibfnamefont {L.}~\bibnamefont {Kronik}}, \ and\ \bibinfo
  {author} {\bibfnamefont {J.~B.}\ \bibnamefont {Neaton}},\ }\href {\doibase
  10.1103/PhysRevB.85.125307} {\bibfield  {journal} {\bibinfo  {journal} {Phys.
  Rev. B}\ }\textbf {\bibinfo {volume} {85}},\ \bibinfo {pages} {125307}
  (\bibinfo {year} {2012}{\natexlab{a}})}\BibitemShut {NoStop}%
\bibitem [{\citenamefont {Sharifzadeh}\ \emph {et~al.}(2013)\citenamefont
  {Sharifzadeh}, \citenamefont {Darancet}, \citenamefont {Kronik},\ and\
  \citenamefont {Neaton}}]{sharifzadeh_jpcl2013}%
  \BibitemOpen
  \bibfield  {author} {\bibinfo {author} {\bibfnamefont {S.}~\bibnamefont
  {Sharifzadeh}}, \bibinfo {author} {\bibfnamefont {P.}~\bibnamefont
  {Darancet}}, \bibinfo {author} {\bibfnamefont {L.}~\bibnamefont {Kronik}}, \
  and\ \bibinfo {author} {\bibfnamefont {J.~B.}\ \bibnamefont {Neaton}},\
  }\href {\doibase 10.1021/jz401069f} {\bibfield  {journal} {\bibinfo
  {journal} {J. Phys. Chem. Lett.}\ }\textbf {\bibinfo {volume} {4}},\ \bibinfo
  {pages} {2197} (\bibinfo {year} {2013})},\ \Eprint
  {http://arxiv.org/abs/http://dx.doi.org/10.1021/jz401069f}
  {http://dx.doi.org/10.1021/jz401069f} \BibitemShut {NoStop}%
\bibitem [{\citenamefont {Rebolini}\ \emph {et~al.}(2014)\citenamefont
  {Rebolini}, \citenamefont {Toulouse}, \citenamefont {Teale}, \citenamefont
  {Helgaker},\ and\ \citenamefont {Savin}}]{rebolini_jcp2014}%
  \BibitemOpen
  \bibfield  {author} {\bibinfo {author} {\bibfnamefont {E.}~\bibnamefont
  {Rebolini}}, \bibinfo {author} {\bibfnamefont {J.}~\bibnamefont {Toulouse}},
  \bibinfo {author} {\bibfnamefont {A.~M.}\ \bibnamefont {Teale}}, \bibinfo
  {author} {\bibfnamefont {T.}~\bibnamefont {Helgaker}}, \ and\ \bibinfo
  {author} {\bibfnamefont {A.}~\bibnamefont {Savin}},\ }\href {\doibase
  http://dx.doi.org/10.1063/1.4890652} {\bibfield  {journal} {\bibinfo
  {journal} {J. Chem. Phys.}\ }\textbf {\bibinfo {volume} {141}},\ \bibinfo
  {eid} {044123} (\bibinfo {year} {2014})}\BibitemShut {NoStop}%
\bibitem [{\citenamefont {Rocca}\ \emph {et~al.}(2014)\citenamefont {Rocca},
  \citenamefont {V{\"o}r{\"o}s}, \citenamefont {Gali},\ and\ \citenamefont
  {Galli}}]{rocca_jctc2014}%
  \BibitemOpen
  \bibfield  {author} {\bibinfo {author} {\bibfnamefont {D.}~\bibnamefont
  {Rocca}}, \bibinfo {author} {\bibfnamefont {M.}~\bibnamefont
  {V{\"o}r{\"o}s}}, \bibinfo {author} {\bibfnamefont {A.}~\bibnamefont {Gali}},
  \ and\ \bibinfo {author} {\bibfnamefont {G.}~\bibnamefont {Galli}},\ }\href
  {\doibase 10.1021/ct5000956} {\bibfield  {journal} {\bibinfo  {journal} {J.
  Chem. Theory Comput.}\ }\textbf {\bibinfo {volume} {10}},\ \bibinfo {pages}
  {3290} (\bibinfo {year} {2014})},\ \Eprint
  {http://arxiv.org/abs/http://dx.doi.org/10.1021/ct5000956}
  {http://dx.doi.org/10.1021/ct5000956} \BibitemShut {NoStop}%
\bibitem [{\citenamefont {Blase}\ and\ \citenamefont
  {Attaccalite}(2011)}]{blase_apl2011}%
  \BibitemOpen
  \bibfield  {author} {\bibinfo {author} {\bibfnamefont {X.}~\bibnamefont
  {Blase}}\ and\ \bibinfo {author} {\bibfnamefont {C.}~\bibnamefont
  {Attaccalite}},\ }\href {\doibase http://dx.doi.org/10.1063/1.3655352}
  {\bibfield  {journal} {\bibinfo  {journal} {Appl. Phys. Lett.}\ }\textbf
  {\bibinfo {volume} {99}},\ \bibinfo {eid} {171909} (\bibinfo {year}
  {2011})}\BibitemShut {NoStop}%
\bibitem [{\citenamefont {Hedin}(1965)}]{hedin_pr1965}%
  \BibitemOpen
  \bibfield  {author} {\bibinfo {author} {\bibfnamefont {L.}~\bibnamefont
  {Hedin}},\ }\href {\doibase 10.1103/PhysRev.139.A796} {\bibfield  {journal}
  {\bibinfo  {journal} {Phys. Rev.}\ }\textbf {\bibinfo {volume} {139}},\
  \bibinfo {pages} {A796} (\bibinfo {year} {1965})}\BibitemShut {NoStop}%
\bibitem [{\citenamefont {Aulbur}, \citenamefont {J\"onsson},\ and\
  \citenamefont {Wilkins}(2000)}]{aulbur_review2000}%
  \BibitemOpen
  \bibfield  {author} {\bibinfo {author} {\bibfnamefont {W.~G.}\ \bibnamefont
  {Aulbur}}, \bibinfo {author} {\bibfnamefont {L.}~\bibnamefont {J\"onsson}}, \
  and\ \bibinfo {author} {\bibfnamefont {J.~W.}\ \bibnamefont {Wilkins}},\
  }\href@noop {} {\bibfield  {journal} {\bibinfo  {journal} {Solid State
  Phys.}\ }\textbf {\bibinfo {volume} {54}},\ \bibinfo {pages} {1} (\bibinfo
  {year} {2000})}\BibitemShut {NoStop}%
\bibitem [{\citenamefont {Parr}\ and\ \citenamefont {Yang}(1989)}]{parr_book}%
  \BibitemOpen
  \bibfield  {author} {\bibinfo {author} {\bibfnamefont {R.~G.}\ \bibnamefont
  {Parr}}\ and\ \bibinfo {author} {\bibfnamefont {W.}~\bibnamefont {Yang}},\
  }\href@noop {} {\emph {\bibinfo {title} {Density-Functional Theory of Atoms
  and Molecules}}}\ (\bibinfo  {publisher} {Oxford University Press},\ \bibinfo
  {address} {New York},\ \bibinfo {year} {1989})\BibitemShut {NoStop}%
\bibitem [{\citenamefont {Cohen}, \citenamefont {Mori-S{\'a}nchez},\ and\
  \citenamefont {Yang}(2008)}]{cohen_science2008}%
  \BibitemOpen
  \bibfield  {author} {\bibinfo {author} {\bibfnamefont {A.~J.}\ \bibnamefont
  {Cohen}}, \bibinfo {author} {\bibfnamefont {P.}~\bibnamefont
  {Mori-S{\'a}nchez}}, \ and\ \bibinfo {author} {\bibfnamefont
  {W.}~\bibnamefont {Yang}},\ }\href {\doibase {10.1126/science.1158722}}
  {\bibfield  {journal} {\bibinfo  {journal} {Science}\ }\textbf {\bibinfo
  {volume} {{321}}},\ \bibinfo {pages} {{792}} (\bibinfo {year}
  {{2008}})}\BibitemShut {NoStop}%
\bibitem [{\citenamefont {Cohen}, \citenamefont {Mori-S\'anchez},\ and\
  \citenamefont {Yang}(2008)}]{cohen_prb2008}%
  \BibitemOpen
  \bibfield  {author} {\bibinfo {author} {\bibfnamefont {A.~J.}\ \bibnamefont
  {Cohen}}, \bibinfo {author} {\bibfnamefont {P.}~\bibnamefont
  {Mori-S\'anchez}}, \ and\ \bibinfo {author} {\bibfnamefont {W.}~\bibnamefont
  {Yang}},\ }\href {\doibase 10.1103/PhysRevB.77.115123} {\bibfield  {journal}
  {\bibinfo  {journal} {Phys. Rev. B}\ }\textbf {\bibinfo {volume} {77}},\
  \bibinfo {pages} {115123} (\bibinfo {year} {2008})}\BibitemShut {NoStop}%
\bibitem [{\citenamefont {Yang}, \citenamefont {Cohen},\ and\ \citenamefont
  {Mori-S{\'a}nchez}(2012)}]{yang_jcp2012}%
  \BibitemOpen
  \bibfield  {author} {\bibinfo {author} {\bibfnamefont {W.}~\bibnamefont
  {Yang}}, \bibinfo {author} {\bibfnamefont {A.~J.}\ \bibnamefont {Cohen}}, \
  and\ \bibinfo {author} {\bibfnamefont {P.}~\bibnamefont {Mori-S{\'a}nchez}},\
  }\href {\doibase http://dx.doi.org/10.1063/1.3702391} {\bibfield  {journal}
  {\bibinfo  {journal} {J. Chem. Phys.}\ }\textbf {\bibinfo {volume} {136}},\
  \bibinfo {eid} {204111} (\bibinfo {year} {2012}),\
  http://dx.doi.org/10.1063/1.3702391}\BibitemShut {NoStop}%
\bibitem [{\citenamefont {Kronik}\ \emph {et~al.}(2012)\citenamefont {Kronik},
  \citenamefont {Stein}, \citenamefont {Refaely-Abramson},\ and\ \citenamefont
  {Baer}}]{kronik_jctc2012}%
  \BibitemOpen
  \bibfield  {author} {\bibinfo {author} {\bibfnamefont {L.}~\bibnamefont
  {Kronik}}, \bibinfo {author} {\bibfnamefont {T.}~\bibnamefont {Stein}},
  \bibinfo {author} {\bibfnamefont {S.}~\bibnamefont {Refaely-Abramson}}, \
  and\ \bibinfo {author} {\bibfnamefont {R.}~\bibnamefont {Baer}},\ }\href
  {\doibase 10.1021/ct2009363} {\bibfield  {journal} {\bibinfo  {journal} {J.
  Chem. Theory Comput.}\ }\textbf {\bibinfo {volume} {8}},\ \bibinfo {pages}
  {1515} (\bibinfo {year} {2012})},\ \Eprint
  {http://arxiv.org/abs/http://dx.doi.org/10.1021/ct2009363}
  {http://dx.doi.org/10.1021/ct2009363} \BibitemShut {NoStop}%
\bibitem [{\citenamefont {Egger}\ \emph {et~al.}(2014)\citenamefont {Egger},
  \citenamefont {Weissman}, \citenamefont {Refaely-Abramson}, \citenamefont
  {Sharifzadeh}, \citenamefont {Dauth}, \citenamefont {Baer}, \citenamefont
  {Kümmel}, \citenamefont {Neaton}, \citenamefont {Zojer},\ and\ \citenamefont
  {Kronik}}]{egger_jctc2014}%
  \BibitemOpen
  \bibfield  {author} {\bibinfo {author} {\bibfnamefont {D.~A.}\ \bibnamefont
  {Egger}}, \bibinfo {author} {\bibfnamefont {S.}~\bibnamefont {Weissman}},
  \bibinfo {author} {\bibfnamefont {S.}~\bibnamefont {Refaely-Abramson}},
  \bibinfo {author} {\bibfnamefont {S.}~\bibnamefont {Sharifzadeh}}, \bibinfo
  {author} {\bibfnamefont {M.}~\bibnamefont {Dauth}}, \bibinfo {author}
  {\bibfnamefont {R.}~\bibnamefont {Baer}}, \bibinfo {author} {\bibfnamefont
  {S.}~\bibnamefont {Kümmel}}, \bibinfo {author} {\bibfnamefont {J.~B.}\
  \bibnamefont {Neaton}}, \bibinfo {author} {\bibfnamefont {E.}~\bibnamefont
  {Zojer}}, \ and\ \bibinfo {author} {\bibfnamefont {L.}~\bibnamefont
  {Kronik}},\ }\href {\doibase 10.1021/ct400956h} {\bibfield  {journal}
  {\bibinfo  {journal} {Journal of Chemical Theory and Computation}\ }\textbf
  {\bibinfo {volume} {10}},\ \bibinfo {pages} {1934} (\bibinfo {year}
  {2014})},\ \Eprint {http://arxiv.org/abs/http://dx.doi.org/10.1021/ct400956h}
  {http://dx.doi.org/10.1021/ct400956h} \BibitemShut {NoStop}%
\bibitem [{\citenamefont {Laurent}\ and\ \citenamefont
  {Jacquemin}(2013)}]{laurent_ijqc2013}%
  \BibitemOpen
  \bibfield  {author} {\bibinfo {author} {\bibfnamefont {A.~D.}\ \bibnamefont
  {Laurent}}\ and\ \bibinfo {author} {\bibfnamefont {D.}~\bibnamefont
  {Jacquemin}},\ }\href {\doibase 10.1002/qua.24438} {\bibfield  {journal}
  {\bibinfo  {journal} {Int. J. Quantum Chem.}\ }\textbf {\bibinfo {volume}
  {113}},\ \bibinfo {pages} {2019} (\bibinfo {year} {2013})}\BibitemShut
  {NoStop}%
\bibitem [{\citenamefont {Schreiber}\ \emph {et~al.}(2008)\citenamefont
  {Schreiber}, \citenamefont {Silva-Junior}, \citenamefont {Sauer},\ and\
  \citenamefont {Thiel}}]{schreiber_jcp2008}%
  \BibitemOpen
  \bibfield  {author} {\bibinfo {author} {\bibfnamefont {M.}~\bibnamefont
  {Schreiber}}, \bibinfo {author} {\bibfnamefont {M.~R.}\ \bibnamefont
  {Silva-Junior}}, \bibinfo {author} {\bibfnamefont {S.~P.~A.}\ \bibnamefont
  {Sauer}}, \ and\ \bibinfo {author} {\bibfnamefont {W.}~\bibnamefont
  {Thiel}},\ }\href {\doibase http://dx.doi.org/10.1063/1.2889385} {\bibfield
  {journal} {\bibinfo  {journal} {J. Chem. Phys.}\ }\textbf {\bibinfo {volume}
  {128}},\ \bibinfo {eid} {134110} (\bibinfo {year} {2008})}\BibitemShut
  {NoStop}%
\bibitem [{\citenamefont {Silva-Junior}\ \emph {et~al.}(2008)\citenamefont
  {Silva-Junior}, \citenamefont {Schreiber}, \citenamefont {Sauer},\ and\
  \citenamefont {Thiel}}]{silvajunior_jcp2008}%
  \BibitemOpen
  \bibfield  {author} {\bibinfo {author} {\bibfnamefont {M.~R.}\ \bibnamefont
  {Silva-Junior}}, \bibinfo {author} {\bibfnamefont {M.}~\bibnamefont
  {Schreiber}}, \bibinfo {author} {\bibfnamefont {S.~P.~A.}\ \bibnamefont
  {Sauer}}, \ and\ \bibinfo {author} {\bibfnamefont {W.}~\bibnamefont
  {Thiel}},\ }\href {\doibase http://dx.doi.org/10.1063/1.2973541} {\bibfield
  {journal} {\bibinfo  {journal} {J. Chem. Phys.}\ }\textbf {\bibinfo {volume}
  {129}},\ \bibinfo {eid} {104103} (\bibinfo {year} {2008})}\BibitemShut
  {NoStop}%
\bibitem [{\citenamefont {Bartlett}\ and\ \citenamefont
  {Musia\l{}}(2007)}]{bartlett_rmp2007}%
  \BibitemOpen
  \bibfield  {author} {\bibinfo {author} {\bibfnamefont {R.~J.}\ \bibnamefont
  {Bartlett}}\ and\ \bibinfo {author} {\bibfnamefont {M.}~\bibnamefont
  {Musia\l{}}},\ }\href {\doibase 10.1103/RevModPhys.79.291} {\bibfield
  {journal} {\bibinfo  {journal} {Rev. Mod. Phys.}\ }\textbf {\bibinfo {volume}
  {79}},\ \bibinfo {pages} {291} (\bibinfo {year} {2007})}\BibitemShut
  {NoStop}%
\bibitem [{\citenamefont {Casida}(1995)}]{casida_book1995}%
  \BibitemOpen
  \bibfield  {author} {\bibinfo {author} {\bibfnamefont {M.~E.}\ \bibnamefont
  {Casida}},\ }\enquote {\bibinfo {title} {Recent advances in density
  functional methods, {P}art {I}},}\ \ (\bibinfo  {publisher} {World
  Scientific},\ \bibinfo {address} {Singapore},\ \bibinfo {year} {1995})\ p.\
  \bibinfo {pages} {155}\BibitemShut {NoStop}%
\bibitem [{\citenamefont {Bruneval}(2012)}]{bruneval_jcp2012}%
  \BibitemOpen
  \bibfield  {author} {\bibinfo {author} {\bibfnamefont {F.}~\bibnamefont
  {Bruneval}},\ }\href {\doibase http://dx.doi.org/10.1063/1.4718428}
  {\bibfield  {journal} {\bibinfo  {journal} {J. Chem. Phys.}\ }\textbf
  {\bibinfo {volume} {136}},\ \bibinfo {eid} {194107} (\bibinfo {year}
  {2012})}\BibitemShut {NoStop}%
\bibitem [{\citenamefont {Grossman}\ \emph {et~al.}(2001)\citenamefont
  {Grossman}, \citenamefont {Rohlfing}, \citenamefont {Mitas}, \citenamefont
  {Louie},\ and\ \citenamefont {Cohen}}]{grossman_prl2001}%
  \BibitemOpen
  \bibfield  {author} {\bibinfo {author} {\bibfnamefont {J.~C.}\ \bibnamefont
  {Grossman}}, \bibinfo {author} {\bibfnamefont {M.}~\bibnamefont {Rohlfing}},
  \bibinfo {author} {\bibfnamefont {L.}~\bibnamefont {Mitas}}, \bibinfo
  {author} {\bibfnamefont {S.~G.}\ \bibnamefont {Louie}}, \ and\ \bibinfo
  {author} {\bibfnamefont {M.~L.}\ \bibnamefont {Cohen}},\ }\href {\doibase
  10.1103/PhysRevLett.86.472} {\bibfield  {journal} {\bibinfo  {journal} {Phys.
  Rev. Lett.}\ }\textbf {\bibinfo {volume} {86}},\ \bibinfo {pages} {472}
  (\bibinfo {year} {2001})}\BibitemShut {NoStop}%
\bibitem [{\citenamefont {Ren}\ \emph {et~al.}(2012)\citenamefont {Ren},
  \citenamefont {Rinke}, \citenamefont {Blum}, \citenamefont {Wieferink},
  \citenamefont {Tkatchenko}, \citenamefont {Sanfilippo}, \citenamefont
  {Reuter},\ and\ \citenamefont {Scheffler}}]{ren_njp2012}%
  \BibitemOpen
  \bibfield  {author} {\bibinfo {author} {\bibfnamefont {X.}~\bibnamefont
  {Ren}}, \bibinfo {author} {\bibfnamefont {P.}~\bibnamefont {Rinke}}, \bibinfo
  {author} {\bibfnamefont {V.}~\bibnamefont {Blum}}, \bibinfo {author}
  {\bibfnamefont {J.}~\bibnamefont {Wieferink}}, \bibinfo {author}
  {\bibfnamefont {A.}~\bibnamefont {Tkatchenko}}, \bibinfo {author}
  {\bibfnamefont {A.}~\bibnamefont {Sanfilippo}}, \bibinfo {author}
  {\bibfnamefont {K.}~\bibnamefont {Reuter}}, \ and\ \bibinfo {author}
  {\bibfnamefont {M.}~\bibnamefont {Scheffler}},\ }\href
  {http://stacks.iop.org/1367-2630/14/i=5/a=053020} {\bibfield  {journal}
  {\bibinfo  {journal} {New Journal of Physics}\ }\textbf {\bibinfo {volume}
  {14}},\ \bibinfo {pages} {053020} (\bibinfo {year} {2012})}\BibitemShut
  {NoStop}%
\bibitem [{\citenamefont {K\"orzd\"orfer}\ and\ \citenamefont
  {Marom}(2012)}]{korzdorfer_prb2012}%
  \BibitemOpen
  \bibfield  {author} {\bibinfo {author} {\bibfnamefont {T.}~\bibnamefont
  {K\"orzd\"orfer}}\ and\ \bibinfo {author} {\bibfnamefont {N.}~\bibnamefont
  {Marom}},\ }\href {\doibase 10.1103/PhysRevB.86.041110} {\bibfield  {journal}
  {\bibinfo  {journal} {Phys. Rev. B}\ }\textbf {\bibinfo {volume} {86}},\
  \bibinfo {pages} {041110} (\bibinfo {year} {2012})}\BibitemShut {NoStop}%
\bibitem [{\citenamefont {Sharifzadeh}\ \emph
  {et~al.}(2012{\natexlab{b}})\citenamefont {Sharifzadeh}, \citenamefont
  {Tamblyn}, \citenamefont {Doak}, \citenamefont {Darancet},\ and\
  \citenamefont {Neaton}}]{sharifzadeh_epjb2012}%
  \BibitemOpen
  \bibfield  {author} {\bibinfo {author} {\bibfnamefont {S.}~\bibnamefont
  {Sharifzadeh}}, \bibinfo {author} {\bibfnamefont {I.}~\bibnamefont
  {Tamblyn}}, \bibinfo {author} {\bibfnamefont {P.}~\bibnamefont {Doak}},
  \bibinfo {author} {\bibfnamefont {P.}~\bibnamefont {Darancet}}, \ and\
  \bibinfo {author} {\bibfnamefont {J.}~\bibnamefont {Neaton}},\ }\href
  {\doibase 10.1140/epjb/e2012-30206-0} {\bibfield  {journal} {\bibinfo
  {journal} {Europ. Phys. J. B}\ }\textbf {\bibinfo {volume} {85}},\ \bibinfo
  {eid} {323} (\bibinfo {year} {2012}{\natexlab{b}}),\
  10.1140/epjb/e2012-30206-0}\BibitemShut {NoStop}%
\bibitem [{\citenamefont {Bruneval}\ and\ \citenamefont
  {Marques}(2013)}]{bruneval_jctc2013}%
  \BibitemOpen
  \bibfield  {author} {\bibinfo {author} {\bibfnamefont {F.}~\bibnamefont
  {Bruneval}}\ and\ \bibinfo {author} {\bibfnamefont {M.~A.~L.}\ \bibnamefont
  {Marques}},\ }\href {\doibase 10.1021/ct300835h} {\bibfield  {journal}
  {\bibinfo  {journal} {J. Chem. Theory Comput.}\ }\textbf {\bibinfo {volume}
  {9}},\ \bibinfo {pages} {324} (\bibinfo {year} {2013})},\ \Eprint
  {http://arxiv.org/abs/http://pubs.acs.org/doi/pdf/10.1021/ct300835h}
  {http://pubs.acs.org/doi/pdf/10.1021/ct300835h} \BibitemShut {NoStop}%
\bibitem [{\citenamefont {Koval}, \citenamefont {Foerster},\ and\ \citenamefont
  {S\'anchez-Portal}(2014)}]{koval_prb2014}%
  \BibitemOpen
  \bibfield  {author} {\bibinfo {author} {\bibfnamefont {P.}~\bibnamefont
  {Koval}}, \bibinfo {author} {\bibfnamefont {D.}~\bibnamefont {Foerster}}, \
  and\ \bibinfo {author} {\bibfnamefont {D.}~\bibnamefont {S\'anchez-Portal}},\
  }\href {\doibase 10.1103/PhysRevB.89.155417} {\bibfield  {journal} {\bibinfo
  {journal} {Phys. Rev. B}\ }\textbf {\bibinfo {volume} {89}},\ \bibinfo
  {pages} {155417} (\bibinfo {year} {2014})}\BibitemShut {NoStop}%
\bibitem [{\citenamefont {K{\"o}rbel}\ \emph {et~al.}(2014)\citenamefont
  {K{\"o}rbel}, \citenamefont {Boulanger}, \citenamefont {Duchemin},
  \citenamefont {Blase}, \citenamefont {Marques},\ and\ \citenamefont
  {Botti}}]{korbel_jctc2014}%
  \BibitemOpen
  \bibfield  {author} {\bibinfo {author} {\bibfnamefont {S.}~\bibnamefont
  {K{\"o}rbel}}, \bibinfo {author} {\bibfnamefont {P.}~\bibnamefont
  {Boulanger}}, \bibinfo {author} {\bibfnamefont {I.}~\bibnamefont {Duchemin}},
  \bibinfo {author} {\bibfnamefont {X.}~\bibnamefont {Blase}}, \bibinfo
  {author} {\bibfnamefont {M.~A.~L.}\ \bibnamefont {Marques}}, \ and\ \bibinfo
  {author} {\bibfnamefont {S.}~\bibnamefont {Botti}},\ }\href {\doibase
  10.1021/ct5003658} {\bibfield  {journal} {\bibinfo  {journal} {J. Chem.
  Theory Comput.}\ }\textbf {\bibinfo {volume} {10}},\ \bibinfo {pages} {3934}
  (\bibinfo {year} {2014})},\ \Eprint
  {http://arxiv.org/abs/http://dx.doi.org/10.1021/ct5003658}
  {http://dx.doi.org/10.1021/ct5003658} \BibitemShut {NoStop}%
\bibitem [{\citenamefont {Ullrich}(2012)}]{ullrich_book2012}%
  \BibitemOpen
  \bibfield  {author} {\bibinfo {author} {\bibfnamefont {C.~A.}\ \bibnamefont
  {Ullrich}},\ }\href@noop {} {\emph {\bibinfo {title} {Time-Dependent
  Density-Functional Theory: Concepts and Applications}}},\ Oxford Graduate
  Texts\ (\bibinfo  {publisher} {Oxford University Press},\ \bibinfo {address}
  {Oxford, New York},\ \bibinfo {year} {2012})\BibitemShut {NoStop}%
\bibitem [{\citenamefont {Shao}\ \emph {et~al.}(2015)\citenamefont {Shao},
  \citenamefont {Felipe H.~{da Jornada}}, \citenamefont {Deslippe},\ and\
  \citenamefont {Louie}}]{shao_arxiv2015}%
  \BibitemOpen
  \bibfield  {author} {\bibinfo {author} {\bibfnamefont {M.}~\bibnamefont
  {Shao}}, \bibinfo {author} {\bibfnamefont {C.~Y.}\ \bibnamefont {Felipe
  H.~{da Jornada}}}, \bibinfo {author} {\bibfnamefont {J.}~\bibnamefont
  {Deslippe}}, \ and\ \bibinfo {author} {\bibfnamefont {S.~G.}\ \bibnamefont
  {Louie}},\ }\href {http://arxiv.org/abs/1501.03830} {\bibfield  {journal}
  {\bibinfo  {journal} {arxiv}\ } (\bibinfo {year} {2015})}\BibitemShut
  {NoStop}%
\bibitem [{\citenamefont {Bruneval}(2015)}]{molgw}%
  \BibitemOpen
  \bibfield  {author} {\bibinfo {author} {\bibfnamefont {F.}~\bibnamefont
  {Bruneval}},\ }\href@noop {} {\enquote {\bibinfo {title} {\textsc{molgw}: a
  slow but accurate many-body perturbation theory code},}\ }\bibinfo
  {howpublished} {https://github.com/bruneval/molgw} (\bibinfo {year}
  {2015})\BibitemShut {NoStop}%
\bibitem [{\citenamefont {Valeev}(2015)}]{libint2}%
  \BibitemOpen
  \bibfield  {author} {\bibinfo {author} {\bibfnamefont {E.~F.}\ \bibnamefont
  {Valeev}},\ }\href@noop {} {\enquote {\bibinfo {title} {A library for the
  evaluation of molecular integrals of many-body operators over gaussian
  functions},}\ }\bibinfo {howpublished} {http://libint.valeyev.net/} (\bibinfo
  {year} {2015})\BibitemShut {NoStop}%
\bibitem [{\citenamefont {Marques}, \citenamefont {Oliveira},\ and\
  \citenamefont {Burnus}(2012)}]{marques_cpc2012}%
  \BibitemOpen
  \bibfield  {author} {\bibinfo {author} {\bibfnamefont {M.~A.~L.}\
  \bibnamefont {Marques}}, \bibinfo {author} {\bibfnamefont {M.~J.~T.}\
  \bibnamefont {Oliveira}}, \ and\ \bibinfo {author} {\bibfnamefont
  {T.}~\bibnamefont {Burnus}},\ }\href {\doibase 10.1016/j.cpc.2012.05.007}
  {\bibfield  {journal} {\bibinfo  {journal} {Chem. Phys. Chem.}\ }\textbf
  {\bibinfo {volume} {183}},\ \bibinfo {pages} {2272} (\bibinfo {year}
  {2012})}\BibitemShut {NoStop}%
\bibitem [{\citenamefont {Koch}\ and\ \citenamefont
  {Jorgensen}(1990)}]{koch_jcp1990}%
  \BibitemOpen
  \bibfield  {author} {\bibinfo {author} {\bibfnamefont {H.}~\bibnamefont
  {Koch}}\ and\ \bibinfo {author} {\bibfnamefont {P.}~\bibnamefont
  {Jorgensen}},\ }\href {\doibase http://dx.doi.org/10.1063/1.458814}
  {\bibfield  {journal} {\bibinfo  {journal} {J. Chem. Phys.}\ }\textbf
  {\bibinfo {volume} {93}},\ \bibinfo {pages} {3333} (\bibinfo {year}
  {1990})}\BibitemShut {NoStop}%
\bibitem [{\citenamefont {Andersson}\ \emph {et~al.}(1990)\citenamefont
  {Andersson}, \citenamefont {Malmqvist}, \citenamefont {Roos}, \citenamefont
  {Sadlej},\ and\ \citenamefont {Wolinski}}]{andersson_jpc1990}%
  \BibitemOpen
  \bibfield  {author} {\bibinfo {author} {\bibfnamefont {K.}~\bibnamefont
  {Andersson}}, \bibinfo {author} {\bibfnamefont {P.~A.}\ \bibnamefont
  {Malmqvist}}, \bibinfo {author} {\bibfnamefont {B.~O.}\ \bibnamefont {Roos}},
  \bibinfo {author} {\bibfnamefont {A.~J.}\ \bibnamefont {Sadlej}}, \ and\
  \bibinfo {author} {\bibfnamefont {K.}~\bibnamefont {Wolinski}},\ }\href
  {\doibase 10.1021/j100377a012} {\bibfield  {journal} {\bibinfo  {journal} {J.
  Phys. Chem.}\ }\textbf {\bibinfo {volume} {94}},\ \bibinfo {pages} {5483}
  (\bibinfo {year} {1990})},\ \Eprint
  {http://arxiv.org/abs/http://dx.doi.org/10.1021/j100377a012}
  {http://dx.doi.org/10.1021/j100377a012} \BibitemShut {NoStop}%
\bibitem [{\citenamefont {Sch{\"a}fer}, \citenamefont {Horn},\ and\
  \citenamefont {Ahlrichs}(1992)}]{schafer_jcp1992}%
  \BibitemOpen
  \bibfield  {author} {\bibinfo {author} {\bibfnamefont {A.}~\bibnamefont
  {Sch{\"a}fer}}, \bibinfo {author} {\bibfnamefont {H.}~\bibnamefont {Horn}}, \
  and\ \bibinfo {author} {\bibfnamefont {R.}~\bibnamefont {Ahlrichs}},\ }\href
  {\doibase http://dx.doi.org/10.1063/1.463096} {\bibfield  {journal} {\bibinfo
   {journal} {J. Chem. Phys.}\ }\textbf {\bibinfo {volume} {97}},\ \bibinfo
  {pages} {2571} (\bibinfo {year} {1992})}\BibitemShut {NoStop}%
\bibitem [{\citenamefont {Kendall}, \citenamefont {Dunning},\ and\
  \citenamefont {Harrison}(1992)}]{kendall_jcp1992}%
  \BibitemOpen
  \bibfield  {author} {\bibinfo {author} {\bibfnamefont {R.~A.}\ \bibnamefont
  {Kendall}}, \bibinfo {author} {\bibfnamefont {T.~H.}\ \bibnamefont
  {Dunning}}, \ and\ \bibinfo {author} {\bibfnamefont {R.~J.}\ \bibnamefont
  {Harrison}},\ }\href {\doibase http://dx.doi.org/10.1063/1.462569} {\bibfield
   {journal} {\bibinfo  {journal} {J. Chem. Phys.}\ }\textbf {\bibinfo {volume}
  {96}},\ \bibinfo {pages} {6796} (\bibinfo {year} {1992})}\BibitemShut
  {NoStop}%
\bibitem [{\citenamefont {Silva-Junior}\ \emph {et~al.}(2010)\citenamefont
  {Silva-Junior}, \citenamefont {Schreiber}, \citenamefont {Sauer},\ and\
  \citenamefont {Thiel}}]{silvajunior_jcp2010}%
  \BibitemOpen
  \bibfield  {author} {\bibinfo {author} {\bibfnamefont {M.~R.}\ \bibnamefont
  {Silva-Junior}}, \bibinfo {author} {\bibfnamefont {M.}~\bibnamefont
  {Schreiber}}, \bibinfo {author} {\bibfnamefont {S.~P.~A.}\ \bibnamefont
  {Sauer}}, \ and\ \bibinfo {author} {\bibfnamefont {W.}~\bibnamefont
  {Thiel}},\ }\href {\doibase http://dx.doi.org/10.1063/1.3499598} {\bibfield
  {journal} {\bibinfo  {journal} {J. Chem. Phys.}\ }\textbf {\bibinfo {volume}
  {133}},\ \bibinfo {eid} {174318} (\bibinfo {year} {2010})}\BibitemShut
  {NoStop}%
\bibitem [{\citenamefont {Rostgaard}, \citenamefont {Jacobsen},\ and\
  \citenamefont {Thygesen}(2010)}]{rostgaard_prb2010}%
  \BibitemOpen
  \bibfield  {author} {\bibinfo {author} {\bibfnamefont {C.}~\bibnamefont
  {Rostgaard}}, \bibinfo {author} {\bibfnamefont {K.~W.}\ \bibnamefont
  {Jacobsen}}, \ and\ \bibinfo {author} {\bibfnamefont {K.~S.}\ \bibnamefont
  {Thygesen}},\ }\href {\doibase 10.1103/PhysRevB.81.085103} {\bibfield
  {journal} {\bibinfo  {journal} {Phys. Rev. B}\ }\textbf {\bibinfo {volume}
  {81}},\ \bibinfo {pages} {085103} (\bibinfo {year} {2010})}\BibitemShut
  {NoStop}%
\bibitem [{\citenamefont {Marom}\ \emph {et~al.}(2012)\citenamefont {Marom},
  \citenamefont {Caruso}, \citenamefont {Ren}, \citenamefont {Hofmann},
  \citenamefont {K\"orzd\"orfer}, \citenamefont {Chelikowsky}, \citenamefont
  {Rubio}, \citenamefont {Scheffler},\ and\ \citenamefont
  {Rinke}}]{marom_prb2012}%
  \BibitemOpen
  \bibfield  {author} {\bibinfo {author} {\bibfnamefont {N.}~\bibnamefont
  {Marom}}, \bibinfo {author} {\bibfnamefont {F.}~\bibnamefont {Caruso}},
  \bibinfo {author} {\bibfnamefont {X.}~\bibnamefont {Ren}}, \bibinfo {author}
  {\bibfnamefont {O.~T.}\ \bibnamefont {Hofmann}}, \bibinfo {author}
  {\bibfnamefont {T.}~\bibnamefont {K\"orzd\"orfer}}, \bibinfo {author}
  {\bibfnamefont {J.~R.}\ \bibnamefont {Chelikowsky}}, \bibinfo {author}
  {\bibfnamefont {A.}~\bibnamefont {Rubio}}, \bibinfo {author} {\bibfnamefont
  {M.}~\bibnamefont {Scheffler}}, \ and\ \bibinfo {author} {\bibfnamefont
  {P.}~\bibnamefont {Rinke}},\ }\href {\doibase 10.1103/PhysRevB.86.245127}
  {\bibfield  {journal} {\bibinfo  {journal} {Phys. Rev. B}\ }\textbf {\bibinfo
  {volume} {86}},\ \bibinfo {pages} {245127} (\bibinfo {year}
  {2012})}\BibitemShut {NoStop}%
\bibitem [{\citenamefont {Caruso}\ \emph {et~al.}(2012)\citenamefont {Caruso},
  \citenamefont {Rinke}, \citenamefont {Ren}, \citenamefont {Scheffler},\ and\
  \citenamefont {Rubio}}]{caruso_prb2012}%
  \BibitemOpen
  \bibfield  {author} {\bibinfo {author} {\bibfnamefont {F.}~\bibnamefont
  {Caruso}}, \bibinfo {author} {\bibfnamefont {P.}~\bibnamefont {Rinke}},
  \bibinfo {author} {\bibfnamefont {X.}~\bibnamefont {Ren}}, \bibinfo {author}
  {\bibfnamefont {M.}~\bibnamefont {Scheffler}}, \ and\ \bibinfo {author}
  {\bibfnamefont {A.}~\bibnamefont {Rubio}},\ }\href {\doibase
  10.1103/PhysRevB.86.081102} {\bibfield  {journal} {\bibinfo  {journal} {Phys.
  Rev. B}\ }\textbf {\bibinfo {volume} {86}},\ \bibinfo {pages} {081102}
  (\bibinfo {year} {2012})}\BibitemShut {NoStop}%
\bibitem [{\citenamefont {Faber}\ \emph {et~al.}(2013)\citenamefont {Faber},
  \citenamefont {Boulanger}, \citenamefont {Duchemin}, \citenamefont
  {Attaccalite},\ and\ \citenamefont {Blase}}]{faber_jcp2013}%
  \BibitemOpen
  \bibfield  {author} {\bibinfo {author} {\bibfnamefont {C.}~\bibnamefont
  {Faber}}, \bibinfo {author} {\bibfnamefont {P.}~\bibnamefont {Boulanger}},
  \bibinfo {author} {\bibfnamefont {I.}~\bibnamefont {Duchemin}}, \bibinfo
  {author} {\bibfnamefont {C.}~\bibnamefont {Attaccalite}}, \ and\ \bibinfo
  {author} {\bibfnamefont {X.}~\bibnamefont {Blase}},\ }\href {\doibase
  http://dx.doi.org/10.1063/1.4830236} {\bibfield  {journal} {\bibinfo
  {journal} {J. Chem. Phys.}\ }\textbf {\bibinfo {volume} {139}},\ \bibinfo
  {eid} {194308} (\bibinfo {year} {2013})}\BibitemShut {NoStop}%
\bibitem [{\citenamefont {Perdew}, \citenamefont {Burke},\ and\ \citenamefont
  {Ernzerhof}(1996)}]{perdew_prl1996}%
  \BibitemOpen
  \bibfield  {author} {\bibinfo {author} {\bibfnamefont {J.~P.}\ \bibnamefont
  {Perdew}}, \bibinfo {author} {\bibfnamefont {K.}~\bibnamefont {Burke}}, \
  and\ \bibinfo {author} {\bibfnamefont {M.}~\bibnamefont {Ernzerhof}},\ }\href
  {\doibase 10.1103/PhysRevLett.77.3865} {\bibfield  {journal} {\bibinfo
  {journal} {Phys. Rev. Lett.}\ }\textbf {\bibinfo {volume} {77}},\ \bibinfo
  {pages} {3865} (\bibinfo {year} {1996})}\BibitemShut {NoStop}%
\bibitem [{\citenamefont {Stephens}\ \emph {et~al.}(1994)\citenamefont
  {Stephens}, \citenamefont {Devlin}, \citenamefont {Chabalowski},\ and\
  \citenamefont {Frisch}}]{stephens_jpc1994}%
  \BibitemOpen
  \bibfield  {author} {\bibinfo {author} {\bibfnamefont {P.~J.}\ \bibnamefont
  {Stephens}}, \bibinfo {author} {\bibfnamefont {F.~J.}\ \bibnamefont
  {Devlin}}, \bibinfo {author} {\bibfnamefont {C.~F.}\ \bibnamefont
  {Chabalowski}}, \ and\ \bibinfo {author} {\bibfnamefont {M.~J.}\ \bibnamefont
  {Frisch}},\ }\href {\doibase 10.1021/j100096a001} {\bibfield  {journal}
  {\bibinfo  {journal} {J. Phys. Chem.}\ }\textbf {\bibinfo {volume} {98}},\
  \bibinfo {pages} {11623} (\bibinfo {year} {1994})},\ \Eprint
  {http://arxiv.org/abs/http://dx.doi.org/10.1021/j100096a001}
  {http://dx.doi.org/10.1021/j100096a001} \BibitemShut {NoStop}%
\bibitem [{\citenamefont {Becke}(1993)}]{becke_jcp1993}%
  \BibitemOpen
  \bibfield  {author} {\bibinfo {author} {\bibfnamefont {A.~D.}\ \bibnamefont
  {Becke}},\ }\href {\doibase http://dx.doi.org/10.1063/1.464304} {\bibfield
  {journal} {\bibinfo  {journal} {J. Chem. Phys.}\ }\textbf {\bibinfo {volume}
  {98}},\ \bibinfo {pages} {1372} (\bibinfo {year} {1993})}\BibitemShut
  {NoStop}%
\bibitem [{\citenamefont {Frisch}\ \emph {et~al.}(2009)\citenamefont {Frisch},
  \citenamefont {Trucks}, \citenamefont {Schlegel}, \citenamefont {Scuseria},
  \citenamefont {Robb}, \citenamefont {Cheeseman}, \citenamefont {Scalmani},
  \citenamefont {Barone}, \citenamefont {Mennucci}, \citenamefont {Petersson},
  \citenamefont {Nakatsuji}, \citenamefont {Caricato}, \citenamefont {Li},
  \citenamefont {Hratchian}, \citenamefont {Izmaylov}, \citenamefont {Bloino},
  \citenamefont {Zheng}, \citenamefont {Sonnenberg}, \citenamefont {Hada},
  \citenamefont {Ehara}, \citenamefont {Toyota}, \citenamefont {Fukuda},
  \citenamefont {Hasegawa}, \citenamefont {Ishida}, \citenamefont {Nakajima},
  \citenamefont {Honda}, \citenamefont {Kitao}, \citenamefont {Nakai},
  \citenamefont {Vreven}, \citenamefont {Montgomery}, \citenamefont {Peralta},
  \citenamefont {Ogliaro}, \citenamefont {Bearpark}, \citenamefont {Heyd},
  \citenamefont {Brothers}, \citenamefont {Kudin}, \citenamefont {Staroverov},
  \citenamefont {Kobayashi}, \citenamefont {Normand}, \citenamefont
  {Raghavachari}, \citenamefont {Rendell}, \citenamefont {Burant},
  \citenamefont {Iyengar}, \citenamefont {Tomasi}, \citenamefont {Cossi},
  \citenamefont {Rega}, \citenamefont {Millam}, \citenamefont {Klene},
  \citenamefont {Knox}, \citenamefont {Cross}, \citenamefont {Bakken},
  \citenamefont {Adamo}, \citenamefont {Jaramillo}, \citenamefont {Gomperts},
  \citenamefont {Stratmann}, \citenamefont {Yazyev}, \citenamefont {Austin},
  \citenamefont {Cammi}, \citenamefont {Pomelli}, \citenamefont {Ochterski},
  \citenamefont {Martin}, \citenamefont {Morokuma}, \citenamefont {Zakrzewski},
  \citenamefont {Voth}, \citenamefont {Salvador}, \citenamefont {Dannenberg},
  \citenamefont {Dapprich}, \citenamefont {Daniels}, \citenamefont {Farkas},
  \citenamefont {Foresman}, \citenamefont {Ortiz}, \citenamefont {Cioslowski},\
  and\ \citenamefont {Fox}}]{gaussian09}%
  \BibitemOpen
  \bibfield  {author} {\bibinfo {author} {\bibfnamefont {M.~J.}\ \bibnamefont
  {Frisch}}, \bibinfo {author} {\bibfnamefont {G.~W.}\ \bibnamefont {Trucks}},
  \bibinfo {author} {\bibfnamefont {H.~B.}\ \bibnamefont {Schlegel}}, \bibinfo
  {author} {\bibfnamefont {G.~E.}\ \bibnamefont {Scuseria}}, \bibinfo {author}
  {\bibfnamefont {M.~A.}\ \bibnamefont {Robb}}, \bibinfo {author}
  {\bibfnamefont {J.~R.}\ \bibnamefont {Cheeseman}}, \bibinfo {author}
  {\bibfnamefont {G.}~\bibnamefont {Scalmani}}, \bibinfo {author}
  {\bibfnamefont {V.}~\bibnamefont {Barone}}, \bibinfo {author} {\bibfnamefont
  {B.}~\bibnamefont {Mennucci}}, \bibinfo {author} {\bibfnamefont {G.~A.}\
  \bibnamefont {Petersson}}, \bibinfo {author} {\bibfnamefont {H.}~\bibnamefont
  {Nakatsuji}}, \bibinfo {author} {\bibfnamefont {M.}~\bibnamefont {Caricato}},
  \bibinfo {author} {\bibfnamefont {X.}~\bibnamefont {Li}}, \bibinfo {author}
  {\bibfnamefont {H.~P.}\ \bibnamefont {Hratchian}}, \bibinfo {author}
  {\bibfnamefont {A.~F.}\ \bibnamefont {Izmaylov}}, \bibinfo {author}
  {\bibfnamefont {J.}~\bibnamefont {Bloino}}, \bibinfo {author} {\bibfnamefont
  {G.}~\bibnamefont {Zheng}}, \bibinfo {author} {\bibfnamefont {J.~L.}\
  \bibnamefont {Sonnenberg}}, \bibinfo {author} {\bibfnamefont
  {M.}~\bibnamefont {Hada}}, \bibinfo {author} {\bibfnamefont {M.}~\bibnamefont
  {Ehara}}, \bibinfo {author} {\bibfnamefont {K.}~\bibnamefont {Toyota}},
  \bibinfo {author} {\bibfnamefont {R.}~\bibnamefont {Fukuda}}, \bibinfo
  {author} {\bibfnamefont {J.}~\bibnamefont {Hasegawa}}, \bibinfo {author}
  {\bibfnamefont {M.}~\bibnamefont {Ishida}}, \bibinfo {author} {\bibfnamefont
  {T.}~\bibnamefont {Nakajima}}, \bibinfo {author} {\bibfnamefont
  {Y.}~\bibnamefont {Honda}}, \bibinfo {author} {\bibfnamefont
  {O.}~\bibnamefont {Kitao}}, \bibinfo {author} {\bibfnamefont
  {H.}~\bibnamefont {Nakai}}, \bibinfo {author} {\bibfnamefont
  {T.}~\bibnamefont {Vreven}}, \bibinfo {author} {\bibfnamefont {J.~A.}\
  \bibnamefont {Montgomery}, \bibfnamefont {{Jr.}}}, \bibinfo {author}
  {\bibfnamefont {J.~E.}\ \bibnamefont {Peralta}}, \bibinfo {author}
  {\bibfnamefont {F.}~\bibnamefont {Ogliaro}}, \bibinfo {author} {\bibfnamefont
  {M.}~\bibnamefont {Bearpark}}, \bibinfo {author} {\bibfnamefont {J.~J.}\
  \bibnamefont {Heyd}}, \bibinfo {author} {\bibfnamefont {E.}~\bibnamefont
  {Brothers}}, \bibinfo {author} {\bibfnamefont {K.~N.}\ \bibnamefont {Kudin}},
  \bibinfo {author} {\bibfnamefont {V.~N.}\ \bibnamefont {Staroverov}},
  \bibinfo {author} {\bibfnamefont {R.}~\bibnamefont {Kobayashi}}, \bibinfo
  {author} {\bibfnamefont {J.}~\bibnamefont {Normand}}, \bibinfo {author}
  {\bibfnamefont {K.}~\bibnamefont {Raghavachari}}, \bibinfo {author}
  {\bibfnamefont {A.}~\bibnamefont {Rendell}}, \bibinfo {author} {\bibfnamefont
  {J.~C.}\ \bibnamefont {Burant}}, \bibinfo {author} {\bibfnamefont {S.~S.}\
  \bibnamefont {Iyengar}}, \bibinfo {author} {\bibfnamefont {J.}~\bibnamefont
  {Tomasi}}, \bibinfo {author} {\bibfnamefont {M.}~\bibnamefont {Cossi}},
  \bibinfo {author} {\bibfnamefont {N.}~\bibnamefont {Rega}}, \bibinfo {author}
  {\bibfnamefont {J.~M.}\ \bibnamefont {Millam}}, \bibinfo {author}
  {\bibfnamefont {M.}~\bibnamefont {Klene}}, \bibinfo {author} {\bibfnamefont
  {J.~E.}\ \bibnamefont {Knox}}, \bibinfo {author} {\bibfnamefont {J.~B.}\
  \bibnamefont {Cross}}, \bibinfo {author} {\bibfnamefont {V.}~\bibnamefont
  {Bakken}}, \bibinfo {author} {\bibfnamefont {C.}~\bibnamefont {Adamo}},
  \bibinfo {author} {\bibfnamefont {J.}~\bibnamefont {Jaramillo}}, \bibinfo
  {author} {\bibfnamefont {R.}~\bibnamefont {Gomperts}}, \bibinfo {author}
  {\bibfnamefont {R.~E.}\ \bibnamefont {Stratmann}}, \bibinfo {author}
  {\bibfnamefont {O.}~\bibnamefont {Yazyev}}, \bibinfo {author} {\bibfnamefont
  {A.~J.}\ \bibnamefont {Austin}}, \bibinfo {author} {\bibfnamefont
  {R.}~\bibnamefont {Cammi}}, \bibinfo {author} {\bibfnamefont
  {C.}~\bibnamefont {Pomelli}}, \bibinfo {author} {\bibfnamefont {J.~W.}\
  \bibnamefont {Ochterski}}, \bibinfo {author} {\bibfnamefont {R.~L.}\
  \bibnamefont {Martin}}, \bibinfo {author} {\bibfnamefont {K.}~\bibnamefont
  {Morokuma}}, \bibinfo {author} {\bibfnamefont {V.~G.}\ \bibnamefont
  {Zakrzewski}}, \bibinfo {author} {\bibfnamefont {G.~A.}\ \bibnamefont
  {Voth}}, \bibinfo {author} {\bibfnamefont {P.}~\bibnamefont {Salvador}},
  \bibinfo {author} {\bibfnamefont {J.~J.}\ \bibnamefont {Dannenberg}},
  \bibinfo {author} {\bibfnamefont {S.}~\bibnamefont {Dapprich}}, \bibinfo
  {author} {\bibfnamefont {A.~D.}\ \bibnamefont {Daniels}}, \bibinfo {author}
  {\bibfnamefont {{\"O}.}~\bibnamefont {Farkas}}, \bibinfo {author}
  {\bibfnamefont {J.~B.}\ \bibnamefont {Foresman}}, \bibinfo {author}
  {\bibfnamefont {J.~V.}\ \bibnamefont {Ortiz}}, \bibinfo {author}
  {\bibfnamefont {J.}~\bibnamefont {Cioslowski}}, \ and\ \bibinfo {author}
  {\bibfnamefont {D.~J.}\ \bibnamefont {Fox}},\ }\href@noop {} {\enquote
  {\bibinfo {title} {Gaussian~09 {R}evision {C}.01},}\ } (\bibinfo {year}
  {2009}),\ \bibinfo {note} {gaussian Inc. Wallingford CT 2009}\BibitemShut
  {NoStop}%
\bibitem [{\citenamefont {Refaely-Abramson}, \citenamefont {Baer},\ and\
  \citenamefont {Kronik}(2011)}]{refaely_prb2011}%
  \BibitemOpen
  \bibfield  {author} {\bibinfo {author} {\bibfnamefont {S.}~\bibnamefont
  {Refaely-Abramson}}, \bibinfo {author} {\bibfnamefont {R.}~\bibnamefont
  {Baer}}, \ and\ \bibinfo {author} {\bibfnamefont {L.}~\bibnamefont
  {Kronik}},\ }\href {\doibase 10.1103/PhysRevB.84.075144} {\bibfield
  {journal} {\bibinfo  {journal} {Phys. Rev. B}\ }\textbf {\bibinfo {volume}
  {84}},\ \bibinfo {pages} {075144} (\bibinfo {year} {2011})}\BibitemShut
  {NoStop}%
\end{thebibliography}

%

\end{document}